%% file: main.tex
\documentclass[fleqn,usenatbib]{mnras}

\usepackage[T1]{fontenc}

\DeclareRobustCommand{\VAN}[3]{#2}
\let\VANthebibliography\thebibliography
\def\thebibliography{\DeclareRobustCommand{\VAN}[3]{##3}\VANthebibliography}

\usepackage{graphicx}	
\usepackage{amsmath}	
\usepackage{amssymb}	
\usepackage{newtxtext,newtxmath}

\title[Cosmic void exclusion models]{Cosmic void exclusion models and their impact on the distance scale measurements from large scale structure}

\author[A. Variu et al.]{
\parbox{\textwidth}{
Andrei Variu,$^{1}$\thanks{E-mail: andrei.variu@epfl.ch}
Cheng Zhao,$^{1}$\thanks{E-mail: cheng.zhao@epfl.ch}
Daniel Forero-Sánchez,$^{1}$
Chia-Hsun Chuang,$^{2}$
Francisco-Shu Kitaura,$^{3,4}$
Charling Tao,$^{5}$
Amélie Tamone,$^{1}$
Jean-Paul Kneib$^{1}$
}
\\
\small $^{1}$Institute of Physics, Laboratory of Astrophysics, École Polytechnique Fédérale de Lausanne (EPFL), Observatoire de Sauverny, CH-1290 Versoix, Switzerland\\
\small $^{2}$Department of Physics and Astronomy, University of Utah, Salt Lake City, UT 84112, USA\\
\small $^{3}$Instituto de Astrofísica de Canarias, s/n, E-38205, La Laguna, Tenerife, Spain\\
\small $^{4}$Departamento de Astrofísica, Universidad de La Laguna, E-38206, La Laguna, Tenerife, Spain\\
\small $^{5}$CPPM, Aix-Marseille Université, CNRS/IN2P3, CPPM UMR 7346, F13288 Marseille, France\\
}

\date{Accepted XXX. Received YYY; in original form ZZZ}

\pubyear{2022}

\begin{document}
\label{firstpage}
\pagerange{\pageref{firstpage}--\pageref{lastpage}}
\maketitle

\begin{abstract}
Baryonic Acoustic Oscillations (BAOs) studies based on the clustering of voids and matter tracers provide important constraints on cosmological parameters related to the expansion of the Universe. However, modelling the void exclusion effect is an important challenge for fully exploiting the potential of this kind of analyses.
We thus develop two numerical methods to describe the clustering of cosmic voids. Neither model requires additional cosmological information beyond that assumed within the galaxy de-wiggled model. The models consist in power spectra whose performance we assess in comparison to a parabolic model on \textsc{Patchy} cubic and light-cone mocks. Moreover, we test their robustness against systematic effects and the reconstruction technique. The void model power spectra and the parabolic model with a fixed parameter provide strongly correlated values for the Alcock-Paczynski ($\alpha$) parameter, for boxes and light-cones likewise. The resulting $\alpha$ values -- for all three models -- are unbiased and their uncertainties are correctly estimated. However, the numerical models show less variation with the fitting range compared to the parabolic one. The Bayesian evidence suggests that the numerical techniques are often favoured compared to the parabolic model.
Moreover, the void model power spectra computed on boxes can describe the void clustering from light-cones as well as from boxes. The same void model power spectra can be used for the study of pre- and post-reconstructed data-sets. Lastly, the two numerical techniques are resilient against the studied systematic effects. Consequently, using either of the two new void models, one can more robustly measure cosmological parameters.
 
\end{abstract}




\input{files/introduction}

\input{files/data}

\input{files/methodology}

\input{files/results}

\input{files/conclusion}

\section*{Acknowledgements}

AV, CZ, DFS, AT acknowledge support from the Swiss National Science Foundation (SNF) "Cosmology with 3D Maps of the Universe" research grant, 200020\_175751 and 200020\_207379. FSK acknowledges the grants SEV-2015-0548, RYC2015-18693, and AYA2017-89891-P. CT is supported by Tsinghua University and sino french CNRS-CAS international laboratories LIA Origins and FCPPL.


\section*{Data Availability}
The \textsc{Patchy} boxes used in this study can be provided upon request to CZ.



\bibliographystyle{mnras}
\bibliography{refs} 




\appendix
\input{files/appendix}


\bsp	
\label{lastpage}
\end{document}

%% file: files/introduction.tex
\section{Introduction}
In order to measure cosmological parameters and better understand the Universe and its expansion, multiple techniques have been developed and implemented; one of them is the study of the Baryonic Acoustic Oscillations (BAOs). They are oscillations in the primordial plasma that have altered the matter distribution in the early Universe, leaving an imprint that has been initially observed in the spectra of Cosmic Microwave Background (CMB) temperature anisotropies \citep[e.g. ][]{2003ApJS..148..135H, 2020A&A...641A...6P}. 

The large spectroscopic surveys provide complementary BAO constraints to CMB. Currently, the most precise BAO studies using the 3D clustering statistics of galaxies have been achieved by Baryon Oscillation Spectroscopic Survey \citep[BOSS;][]{2017MNRAS.470.2617A} and extended-BOSS \citep[eBOSS;][]{2021PhRvD.103h3533A}.
The ongoing Dark Energy Spectroscopic Instrument \citep[DESI;][]{DESI2016} plans to further improve the precision of the BAO measurements by increasing the number density of tracers and mapping larger volumes. Meanwhile, the future Cosmology Redshift Survey \citep[CRS;][]{2019Msngr.175...50R}, part of 4-metre Multi-Object Spectroscopic Telescope \citep[4MOST;][]{deJong2019} survey, will provide complementary measurements to DESI by scanning different regions on the sky. In addition to the clustering of galaxies -- e.g. luminous red galaxies \citep[LRG;][]{2017MNRAS.464.1168R, 2017MNRAS.464.3409B}, emission line galaxies \citep[ELG;][]{10.1093/mnras/staa3336} -- the BAO feature has been detected in the clustering of quasi-stellar objects \citep[QSO;][]{10.1093/mnras/stx2630}, Lyman $\alpha$ forests \citep[Ly$\alpha$ forests;][]{refId0} and cosmic voids \citep{PhysRevLett.116.171301}.

While the matter tracers -- except Ly$\alpha$ forests -- are directly observable, the cosmic voids are detected from the positions of the former. In general, cosmic voids are regions in space emptied of luminous objects that trace the under-dense zones of the density field \citep[see review of ][]{Weygaert2011}. However, in practice, there are multiple definitions and thus different algorithms to detect them \citep[e.g.][and references therein]{2005MNRAS.363..977P, 2007MNRAS.380..551P, 2008MNRAS.386.2101N, 2015A&C.....9....1S,10.1093/mnras/stw660}. This allows for a greater diversity of cosmological measurements. For example,
cosmic voids are part of BAO studies \citep[e.g][]{ImproveBAOvoids_BOSS, 2021PhRvD.103d3502C, 10.1093/mnras/stac390}, their geometry is involved in performing Alcock-Paczynski tests \citep[e.g.][]{2012ApJ...761..187S, 2017ApJ...835..160M}, their cross-clustering with galaxies has been used in Redshift-Space-Distortions (RSD) studies \citep[e.g.][]{2016PhRvL.117i1302H, 2019PhRvD.100b3504N, 2020JCAP...12..023H, 2022MNRAS.509.1871C}.

Multi-tracer analyses \citep{ImproveBAOvoids_BOSS, 10.1093/mnras/stac390} of galaxies with voids determined using the  Delaunay trIangulation Void findEr \citep[\textsc{DIVE};][]{10.1093/mnras/stw660} -- code that uses the Delaunay Triangulation \citep[DT;][]{Delaunay1934} on the positions of the matter tracers -- show improvements on the precision of Alcock--Paczynski parameter \citep[$\alpha$;][]{1979Natur.281..358A} of the order of 10 per cent compared to galaxy-only measurements. However, these studies imply the additional challenge of modelling the void clustering. Compared to the matter tracers, voids have large sizes, hence their exclusion has a stronger impact on the clustering \citep{2014PhRvL.112d1304H}. In consequence, \citet{ImproveBAOvoids_BOSS} have developed a more general model than the galaxy de-wiggled one \citep{Xu_2012MNRAS.427.2146X} in order to correctly account for this difference.

The purpose of this paper is to introduce two numerical methods that can be used in the modified de-wiggled model to provide a description of the void exclusion effect. The principle behind the two methods is to first create a halo catalogue by assigning them directly on the density field corresponding to the initial conditions and then detect the voids. Finally, the computed void power spectrum represents the model for the void exclusion.

Section~\ref{sec:data} presents the simulations involved in assessing the performance of the void model power spectra. The description of the two numerical techniques and the methodology employed in testing them are described in Section~\ref{sec:methodology}. Section~\ref{sec:results} shows the results of the performance and robustness tests that have been effectuated on the numerical techniques. The last section concludes the current article.

%% file: files/data.tex
\section{Data}
\label{sec:data}
\subsection{\textsc{Patchy} boxes}
In this study, we use two sets of $2.5\,h^{-1}\mathrm{Gpc}$ cubic mock catalogues obtained using the PerturbAtion Theory Catalogue generator of Halo and galaxY distributions \citep[\textsc{Patchy;}][]{10.1093/mnrasl/slt172}. This generator uses the Augmented Lagrangian Perturbation Theory \citep[\textsc{ALPT;}][]{10.1093/mnrasl/slt101} to model the structure formation and then it assigns biased tracers (e.g. haloes or galaxies) to the density field based on a bias model. 

Both sets of \textsc{Patchy} boxes are calibrated against the BigMultiDark (\textsc{BigMD}) $N$-body simulation \citep{10.1093/mnras/stw248}. However, the set of 1000 boxes is tuned to match a \textsc{BigMD} Sub-Halo Abundance Matching (SHAM) galaxy catalogue, whereas the set of 100 mocks is calibrated with a \textsc{BigMD} halo catalogue. 

The reference \textsc{BigMD} dark-matter box has a side length of $2.5\,h^{-1}\mathrm{Gpc}$ and contains $3840^3$ dark-matter particles with a mass of $2.359\times10^{10}\,h^{-1}\mathrm{M}_\odot$ each. The cosmology of the simulation is described by $h = 0.6777$, $\Omega_\Lambda=0.692885$, $\Omega_\mathrm{m}=0.307115$, $\Omega_\mathrm{b}=0.048206$, $n=0.96$, $\sigma_8=0.8228$ \footnote{\url{https://www.cosmosim.org/cms/simulations/bigmdpl/}}.

On one hand, the \textsc{BigMD} SHAM mock is based on the dark-matter snapshot at redshift $z=0.4656$ and has a galaxy density of $n=3.976980 \times 10^{-4}\,h^3\mathrm{Mpc}^{-3}$. On the other hand, the \textsc{BigMD} halo catalogue uses the snapshot at $z=0.5618$ and has a number density of $n=3.5 \times 10^{-4}\,h^3\mathrm{Mpc}^{-3}$.   

\subsection{\textsc{Patchy} light-cones}
\label{sec:data_lc}
\begin{figure}
 \includegraphics[width=\columnwidth]{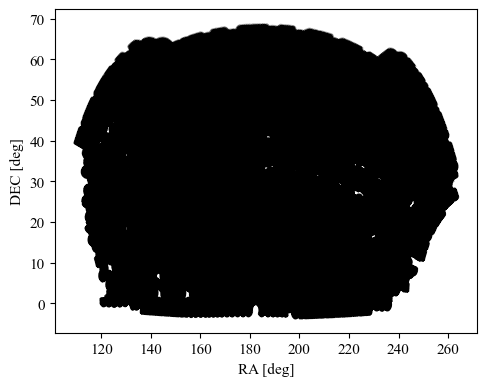}
 \caption{The NGC footprint of the BOSS DR12 \citep{2015ApJS..219...12A} used to build the \textsc{Patchy} light-cones.}
 \label{fig:radec}
\end{figure}

\begin{figure}
 \includegraphics[width=\columnwidth]{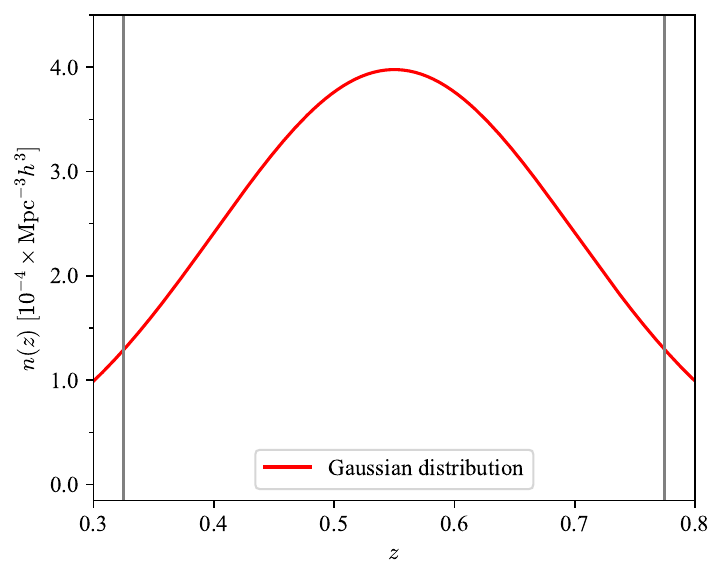}
 \caption{The theoretical radial selection function used to build the \textsc{Patchy} light-cones. The used redshift range is $z\in[0.325, 0.775]$, between the two vertical grey lines.}
 \label{fig:n_z}
\end{figure}

In order to validate the suitability of the numerical models for survey-like data, we construct the Light-Cones (LC) of all the 1000 \textsc{Patchy} galaxy boxes using the \textsc{make\_survey}\footnote{\url{https://github.com/mockFactory/make\_survey}} \citep{10.1093/mnras/stt2071} code. This implies:
\begin{itemize}
    \item the conversion of the $(X, Y, Z)$ euclidean coordinates to Right Ascension (RA), Declination (DEC) and redshift $z$;
    \item the cut of a survey geometry in (RA, DEC);
    \item the application of a radial selection function to sample tracers along the line-of-sight.
\end{itemize}
On one hand, the applied footprint (Figure~\ref{fig:radec}) corresponds to the BOSS DR12\footnote{\url{https://data.sdss.org/sas/dr12/boss/lss/}} Northern-Galactic Cap (NGC) footprint \citep{2015ApJS..219...12A}.
On the other hand, a Gaussian distribution (Figure~\ref{fig:n_z}) is used as a radial selection function, for $z\in[0.325, 0.775]$. This distribution is realistic enough for the current purpose and it allows for the flexibility of choosing the redshift range and the shape.

%% file: files/methodology.tex
\section{Methodology}
\label{sec:methodology}
\subsection{BAO reconstruction}
The BAO reconstruction technique \citep{2007ApJ...664..675E} is used to increase the BAO signal (from the clustering of matter tracers) and thus improve constraints on the cosmological parameters \citep[e.g.][]{anderson_clustering_2014, 2017MNRAS.470.2617A, 10.1093/mnras/staa2800, 10.1093/mnras/staa3336,2021PhRvD.103h3533A}.

The principle of this technique is to estimate the displacement of the biased matter tracers and then move them at positions corresponding to higher redshifts to linearise the density field.
By implementation, this method affects the distribution and the clustering of the matter tracers, thus the distribution of the determined voids and their clustering also change. Given that the reconstruction has been used in multi-tracer analysis of voids and galaxies \citep{ImproveBAOvoids_BOSS,10.1093/mnras/stac390} and it changes the void clustering, it is imperative to test whether the numerical models can describe the voids obtained from reconstructed \textsc{Patchy} mock catalogues.

In the current study, we adopt the iterative method proposed by \citet{10.1093/mnras/stv1581} to perform the reconstruction. 
In practice, we use the code \textsc{Revolver}\footnote{\url{https://github.com/seshnadathur/Revolver}} described in \citet{2019PhRvD.100b3504N}. The required input parameters of the code are the number of iterations (three, in this study) the linear bias of the mock tracers $b=2.2$, the growth rate $f=0.743$ (corresponding to an effective redshift of the simulation boxes $z=0.4656$), the smoothing scale $S=15\,h^{-1}\mathrm{Mpc}$ and the grid size of $512^3$ on which the density field is approximated using a Cloud-In-Cell \citep[CIC;][]{10.1093/mnras/stw1229} mass assignment scheme.

\subsection{Void detection}
We apply the \textsc{DIVE}\footnote{\url{https://github.com/cheng-zhao/DIVE}} code \citep{10.1093/mnras/stw660} to the galaxy and halo catalogues to obtain the DT spheres. Similarly to other methods \citep[e.g][]{2004MNRAS.350..517S,2014PhRvL.112y1302H},
\citet{10.1093/mnras/stw660} have shown that while the small DT spheres are mostly \textit{voids-in-clouds} and have positive matter density contrast, the larger DT spheres (DT voids) are more probably \textit{voids-in-voids} and exhibit a negative matter density contrast. Consequently, a radius based selection -- which depends on the matter tracers' number density -- can discriminate the true tracers of under-dense regions from the possible tracers of over-dense regions. Moreover, \citet{10.1093/mnras/stw884} have proved that a radius based selection can be used to maximise the signal-to-noise ratio of the BAO signal from the clustering of DT voids.

In this study, we are interested in modelling only the DT voids as they have been used in multi-tracer analyses such as \citet{ImproveBAOvoids_BOSS, 10.1093/mnras/stac390} to improve the precision of BAO measurements. Thus, we select the DT spheres with a radius $\mathrm{R}_v\geq16\,h^{-1}\mathrm{Mpc}$ to form the DT void sample. This radius cut is chosen by analogy to \citet{ImproveBAOvoids_BOSS} and based on the studies of \citet{10.1093/mnras/stw884, 2022MNRAS.513.5407F}.
\citet{2022MNRAS.513.5407F} have shown that the void selection based on a constant radius cut yields unbiased BAO measurements when reconstruction is applied on the galaxy catalogue or when systematical effects -- such as a small sample incompleteness -- are present. 
Lastly, \citet{10.1093/mnras/stw660} have observed that by selecting the large DT spheres, the resulting DT void sample has a negative bias, consistently with the detailed results of \citet{2014PhRvL.112d1304H}.

\subsection{Clustering computation}
\subsubsection{Two point correlation function}
In order to compute the 2PCF we use the Fast Correlation Function Calculator\footnote{\url{https://github.com/cheng-zhao/FCFC}} (FCFC) code \citep{2023arXiv230112557Z}, which accepts as input both boxes and light-cones and can employ any type of estimator. In the current study, several estimators have been necessary to correctly account for the specificity of the data sets.

\begin{itemize}
    \item The natural estimator \citep{1974ApJS...28...19P} is used to compute the void auto-2PCF and void-galaxy cross-2PCF from pre-reconstructed boxes and the void auto-2PCF from post-reconstructed boxes:

\begin{equation}
    \xi(s)=\frac{\mathrm{D}_\mathrm{v}\mathrm{D}_\mathrm{v}(s)}{\mathrm{R}_\mathrm{v}\mathrm{R}_\mathrm{v}(s)} - 1,
\end{equation}

\begin{equation}
    \xi(s)=\frac{\mathrm{D}_\mathrm{g}\mathrm{D}_\mathrm{v}(s)}{\mathrm{R}_\mathrm{g}\mathrm{R}_\mathrm{v}(s)} - 1.
\end{equation}

    \item The Landy--Szalay estimator \citep{1993ApJ...412...64L} is needed to compute the void auto-2PCF and void-galaxy cross-2PCF for the light-cones:
\begin{equation}
    \xi(s)=\frac{\mathrm{D}_\mathrm{v}\mathrm{D}_\mathrm{v}(s)-\mathrm{2D}_\mathrm{v}\mathrm{R}_\mathrm{v}(s) + \mathrm{R}_\mathrm{v}\mathrm{R}_\mathrm{v}(s)}{\mathrm{R}_\mathrm{v}\mathrm{R}_\mathrm{v}(s)},
\end{equation}

\begin{equation}
    \xi(s)=\frac{\mathrm{D}_\mathrm{g}\mathrm{D}_\mathrm{v}(s)-\mathrm{R}_\mathrm{g}\mathrm{D}_\mathrm{v}(s)-\mathrm{D}_\mathrm{g}\mathrm{R}_\mathrm{v}(s) + \mathrm{R}_\mathrm{g}\mathrm{R}_\mathrm{v}(s)}{\mathrm{R}_\mathrm{g}\mathrm{R}_\mathrm{v}(s)}.
\end{equation}

    \item A modified version of the Landy--Szalay estimator \citep{2012MNRAS.427.2132P} -- inspired from \citet{1997astro.ph..4241S} -- is required to compute the void-galaxy cross-2PCF from the post-reconstructed boxes:
\begin{equation}
    \xi(s)=\frac{\mathrm{D}_\mathrm{g}\mathrm{D}_\mathrm{v}(s)-\mathrm{S}_\mathrm{g}\mathrm{D}_\mathrm{v}(s)-\mathrm{D}_\mathrm{g}\mathrm{R}_\mathrm{v}(s) + \mathrm{S}_\mathrm{g}\mathrm{R}_\mathrm{v}(s)}{\mathrm{R}_\mathrm{g}\mathrm{R}_\mathrm{v}(s)}.
\end{equation}

\end{itemize}

On one hand, the letter D denotes the data catalogue of voids (D$_\mathrm{v}$) or galaxies (D$_\mathrm{g}$) and thus DD represents the data-data normalised pair counts. On the other hand, the random catalogue is expressed through the letter R that can be related to both voids (R$_\mathrm{v}$) and galaxies (R$_\mathrm{g}$). Consequently, RR and DR serve as the symbols for the random-random and data-random normalised pair counts, respectively. Lastly, S$_\mathrm{g}$ is referring to a galaxy random catalogue that was shifted by the same displacement field as the reconstructed galaxy catalogue and thus $\mathrm{S}_\mathrm{g}\mathrm{R}_\mathrm{v}$ represents the random-random pair counts.

The data-data pair counts can be directly computed given the measured data catalogue. However, in order to compute the data-random and random-random pair counts, one has to construct the random part. 
For boxes, which implicitly have periodic boundary conditions, the RR term (R$_\mathrm{v}$R$_\mathrm{v}$; R$_\mathrm{g}$R$_\mathrm{v}$) can be computed analytically:
\begin{equation}
    \mathrm{RR}(s)=\frac{4\pi(s_2^3 - s_1^3)}{3} \frac{1}{V},
\end{equation}
where $s_2$ and $s_1$ are the boundaries of a separation bin ($s_2 > s_1$) and $s=(s_2+s_1)/2$ for linearly separated bins. 

In contrast, for light-cones, the RR term has to be evaluated on random catalogues which must include the same observational effects as the data catalogues. For galaxies, we initially create a random box (RB) of the same size as the \textsc{BigMD} and \textsc{Patchy} boxes, but ten times denser, by randomly sampling Cartesian positions. Afterwards, we apply \textsc{make\_survey} with the same configurations as for the \textsc{Patchy} boxes in order to obtain a random LC that is ten times denser than the \textsc{Patchy} LC. 

In the case of voids, we adopt a modified version of the 'shuffling' technique \citep{10.1093/mnras/stw884}. \citet{de_Mattia_2019, 2021MNRAS.503.1149Z} have shown that it is necessary to avoid having identical angular and radial positions of objects in the data and the random catalogues, otherwise, the measured clustering is affected. Consequently, to diminish this effect, we stack 100 void \textsc{Patchy} LC mocks. Furthermore, we shuffle the RA-DEC pairs in bins of redshift and void radius. This shuffling maintains the angular coverage, but breaks the correlation between the redshift-radius pair and the RA-DEC pair. Finally, we uniformly and randomly down-sample the resulting shuffled catalogue down to 20 times the void density of the \textsc{Patchy} LC. Having the void and galaxy random catalogues, one can compute the R$_\mathrm{v}$R$_\mathrm{v}$,  R$_\mathrm{g}$R$_\mathrm{v}$, D$_\mathrm{v}$R$_\mathrm{v}$, R$_\mathrm{g}$D$_\mathrm{v}$, D$_\mathrm{g}$R$_\mathrm{v}$  pair counts for LC.

The shifted galaxy random cubic catalogues S$_\mathrm{g}$ are computed during the reconstruction of the \textsc{Patchy} boxes by applying the displacement field that is estimated from the \textsc{Patchy} boxes on the random box RB. This creates a dedicated random catalogue to each of the \textsc{Patchy} boxes. In comparison with galaxies, the void random box is simply constructed by randomly and uniformly sampling Cartesian positions inside a box of side-length of $2500\,h^{-1}\mathrm{Mpc}$, so that the density is ten times larger than the DT void sample. 

We finally compute the pair counts and the 2PCF using 40 separation bins between 0 and 200 $h^{-1}\mathrm{Mpc}$ (i.e. a bin width of $5\,h^{-1}\mathrm{Mpc}$).

\subsubsection{Power spectrum}
\label{sec:power_spectrum}
In the current study, we exploit the \textsc{POWSPEC}\footnote{\url{https://github.com/cheng-zhao/powspec}} code  -- described in \citet{2021MNRAS.503.1149Z} -- to calculate the required power spectra. The density field is estimated using the Cloud-In-Cell \citep[CIC;][]{10.1093/mnras/stw1229} particle assignment scheme and power spectra are computed in $k$ bins of size $0.0025\,h\,\mathrm{Mpc}^{-1}$.

The smoothness of the 2PCFs obtained through the Hankel transform (see Section~\ref{sec:baomodel}) of power spectra depends on the range spanned by the wavenumber $k$ and on the number of power spectra realisations. The large value of $k$ is required to ameliorate the effect of the undulatory shape of the 0-order spherical Bessel function used in the Hankel transform, while the large number of realisations is needed to decrease the noise coming from cosmic variance.
In order to achieve a large enough $k$ interval, we use a grid size of $2048^3$ to measure the power spectra. This provides a $k_\mathrm{max}\sim2.57\, h\,\mathrm{Mpc}^{-1}$ for boxes and a $k_\mathrm{max}\sim1.88\, h\,\mathrm{Mpc}^{-1}$ for light-cones.

Given the fact that we need a large number of realisations to reduce variances, it is computationally-expensive to always use a grid size of $2048^3$. Thus, we also calculate power spectrum realisations using a grid size of $512^3$ in order to have a smoother power spectrum for lower wavenumbers (see Section~\ref{sec:templatecreation} for more details). In this case, we use the grid interlacing technique \citep{10.1093/mnras/stw1229} to reduce the alias effects introduced by the particle assignments scheme.

\subsection{BAO fitting}
\subsubsection{BAO models}
\label{sec:baomodel}

\begin{table}
    \centering
    \begin{tabular}{l|l}
        \hline
            Abbreviation & Description \\
        \hline
            DW      & de-wiggled model, Eq.~\eqref{eq:templatevoids} \\
            PAR     & parabolic model, Eq.~\eqref{eq:parabolic_model} \\
            PAR$_\mathrm{U}$     & PAR with uniform prior, Eq.~\eqref{eq:parabolic_uniform_prior} \\
            PAR$_\mathrm{G}$     & PAR with a prior defined by Eq.~\eqref{eq:par_g_prior_c} \\
            fix c     & PAR with a fixed c parameter,\\
                      & determined from the fit of the average \\
                      & 2PCF from 500 or 1000 realisations\\
            
            SK      & SICKLE, details in Sec.~\ref{sec:sickle} and Tab.~\ref{tab:cg_sk_params} \\
            SK$_\mathrm{B}$      & calibrated SK model based on Boxes having\\
                                 & the same halo number density as the reference\\
            SK$_\mathrm{def}$    & defective SK model, see Tab.~\ref{tab:cg_sk_params} \\
            SK$_\mathrm{LC}$     & the model obtained by applying the survey \\
                                 & geometry (Light-Cone) of the reference on the \\
                                 & halo boxes corresponding to SK$_\mathrm{B}$\\

            CG      & CosmoGAME, details in Sec.~\ref{sec:cosmogame} and Tab.~\ref{tab:cg_sk_params}\\
            CG$_\mathrm{B}$      & same as SK$_\mathrm{B}$ but for CG\\
            CG$_\mathrm{def}$    & same as SK$_\mathrm{def}$ but for CG \\
            CG$_\mathrm{LC}$     & same as SK$_\mathrm{LC}$ but for CG \\
            CG$_\mathrm{80}$     & calibrated CG model based on boxes having \\
                                 & a $20\%$ lower halo number density than the reference \\
            CG$_\mathrm{120}$    & calibrated CG model based on boxes having\\
                                 & a $20\%$ higher halo number density than the reference \\
        
        gv  & void-halo (galaxy) cross-clustering \\
        vv  & void auto-clustering \\
        \hline
    \end{tabular}
    \caption{The abbreviations of the studied models.}
    \label{tab:model_abbreviations}
\end{table}

The theoretical model used to fit the 2PCF is defined as follows \citep{Xu_2012MNRAS.427.2146X}:
\begin{equation}
\label{eq:ximodel}
    \xi_\mathrm{model}(s)\equiv B^2 \xi_\mathrm{t}(\alpha s) + A(s),
\end{equation}
where $B$ tunes the amplitude of the model, $\alpha$ is the Alcock--Paczynski \citep{1979Natur.281..358A} parameter that is related to the position of the BAO peak and $A(s)$ is a function required to describe the broad-band shape of the correlation function, which consists of three nuisance parameters $a_0$, $a_1$, $a_2$:
\begin{equation}
    A(s) = a_0 + a_1 s^{-1} + a_2 s^{-2}.
\end{equation}
\citet{Xu_2012MNRAS.427.2146X} and \citet{10.1093/mnras/stu1681} have shown that this function does not bias the measurement of $\alpha$. Lastly, $\xi_\mathrm{t}$ is the Hankel transform of the template power spectrum $P_\mathrm{t}(k)$ as described in \citet{Xu_2012MNRAS.427.2146X}:
\begin{equation}
\label{eq:dampfouriertransform}
\xi_\mathrm{t}(s)=\int \frac{k^2\mathrm{d}k}{2\pi^2}P_\mathrm{t}(k)j_0(ks)\mathrm{e}^{-k^2a^2}\mathrm{,}
\end{equation}
where $j_0$ is the 0-order spherical Bessel function of the first kind (i.e. the sinc function) and $a=2\,h^{-1}~\mathrm{Mpc}$ is a factor for the Gaussian damping of the Bessel function's wiggles at high-$k$. A more detailed discussion on how the value of $a$ was chosen is presented in Section~\ref{sec:templatecreation}.

In the case of galaxies, the template power spectrum can be expressed by the typical de-wiggled (DW) model \citep{anderson_clustering_2014}:
\begin{equation}
\label{eq:templatevoids}
P_\mathrm{t, DW}(k)=[P_{\text{lin}}(k)-P_{\text{lin,nw}}(k)]\mathrm{e}^{-k^2\Sigma^2_{\text{nl}}/2}+P_{\text{lin,nw}}(k) \mathrm{,}
\end{equation}
where $P_\mathrm{lin}(k)$ is the linear power spectrum that can be obtained using \textsc{CAMB}\footnote{\url{https://camb.info/}} software \citep{Lewis_2000}, $P_\mathrm{lin,nw}(k)$ is the linear power spectrum without the BAO feature (no wiggles, nw) computed using the formula of \citet{eisenstein_baryonic_1998}, and $\Sigma_\mathrm{nl}$ is the damping parameter for BAO \citep{eisenstein_robustness_2007}. In this work, we use the input power spectrum employed in the generation of the \textsc{Patchy} mocks as $P_\mathrm{lin}(k)$ for BAO fittings. This provides a predictable $\alpha$ value in the absence of any systematic effects (see Section~\ref{sec:tension_param} for a discussion on this topic). 

\citet{ImproveBAOvoids_BOSS} have shown that the de-wiggled model is not suitable for voids due to the improper accounting of the broadband shape. More precisely, the exclusion effect of voids \citep{2014PhRvL.112d1304H} affects significantly the clustering and thus the shapes of the 2PCF and power spectrum.

Consequently, \citet{ImproveBAOvoids_BOSS} have introduced a more general template power spectrum that accounts for the exclusion effect:
\begin{equation}
\label{eq:templatepk}
P_\mathrm{t}(k)=\varphi(k) P_\mathrm{t, DW}(k)
\end{equation}
and
\begin{equation}
    \label{eq:additionalfactor}
    \varphi(k)=\frac{P_{\mathrm{t},\mathrm{nw}}(k)}{P_\mathrm{lin,nw}(k)}\mathrm{,}
\end{equation}
where $P_\mathrm{t, nw}(k)$ is the non-wiggled tracer power spectrum, that can practically include the void exclusion effect.

In this paper, we study different methods to model the additional factor introduced in the template power spectrum, whose names and abbreviations are summarised in Table~\ref{tab:model_abbreviations}. The first method is introduced by \citet{ImproveBAOvoids_BOSS} and it consists in approximating the factor with a parabola (parabolic model). The other two methods provide numerical models for the $P_{\mathrm{t},\mathrm{nw}}(k)$ term in three steps:
\begin{enumerate}
    \item create a halo catalogue using gauSsIan moCK tempLate gEnerator (\textsc{SICKLE}\footnote{\url{https://github.com/Andrei-EPFL/SICKLE}}) or Cosmological GAussian Mock gEnerator (\textsc{CosmoGAME}\footnote{\url{https://github.com/cheng-zhao/CosmoGAME}});
    \item apply \textsc{DIVE} on the constructed halo catalogues to get the DT voids;
    \item measure the power spectra of the resulting DT void catalogues.
\end{enumerate}
\textsc{SICKLE} and \textsc{CosmoGAME} are two C codes that:
\begin{enumerate}
    \item generate Gaussian random fields based on $P_\mathrm{lin,nw}(k)$, using the fixed amplitude \citep{2016MNRAS.462L...1A} presented in \citet{10.1093/mnras/stz1233}, in order to decrease the sample variance of halo--halo and halo--void clustering;
    \item assign haloes directly on the Gaussian fields without gravitational evolution.
\end{enumerate}
Nonetheless, the two techniques differ in their halo assignment schemes.

By construction, our methods have the advantage of being generalisable for multiple definitions of voids as one needs to simply apply the required necessary void finder on the resulting \textsc{SICKLE} or \textsc{CosmoGAME} halo catalogue. However, the disadvantage is that they are computationally expensive compared to analytical models. Consequently, we may consider in future studies analytical models based on the pioneering work to model the void exclusion \citep{2014PhRvL.112d1304H} by \citet{2014PhRvD..90j3521C}.

\paragraph{Parabolic model}
\citet{ImproveBAOvoids_BOSS} have shown that the additional factor -- $\varphi(k)$, Eq.~\eqref{eq:additionalfactor} -- can be approximated by a parabola (PAR):
\begin{equation}
\label{eq:parabolic_model}
\varphi(k) \sim 1 + c k^2,
\end{equation}
where $c$ is a free parameter, determined through the fitting process.
In practice, when we fit the 2PCF, we force $c$ to take values only inside a prior interval with a given probability distribution. More details about the prior distribution are discussed in Section~\ref{sec:param_priors}. 

\paragraph{SICKLE}
\label{sec:sickle}

The code generates a Gaussian random field in Fourier space on a grid whose size can be tuned ($N_\mathrm{grid}$). The field is then scaled by a factor $\gamma$ to encode the information about the linear growth and the bias parameter. The resulting field is in an approximation of the matter overdensity field in Fourier space $\Tilde{\delta}(\textbf{k})$. 
Furthermore, $\Tilde{\delta}(\textbf{k})$ is transformed to real space into $\delta_\mathrm{m}(\textbf{r})$ using the implementation of the Discrete Fourier Transform in the \textsc{FFTW}\footnote{\url{http://fftw.org/}} package.

Starting from the matter overdensity field $\delta_\mathrm{m}(\textbf{r})$, haloes are selected by an iterative algorithm inspired from the CIC mass assignment scheme until the desired number of haloes is reached:
\begin{enumerate}
    \item obtain the $(x,y,z)$ position of the maximum overdensity value;
    \item scatter the $(x,y,z)$ position using displacements sampled from a Triangular distribution ($\mathcal{T}(x)=\mathrm{max}(1-|x|,0)$; given by the weight of the CIC scheme) to get a new $(x',y',z')$ position;
    \item assign a halo at $(x',y',z')$;
    \item compute the contribution of the assigned halo to the matter density field using the CIC scheme;
    \item subtract the previously computed contribution from the density field in order to emulate the exclusion of massive haloes;
    \item go to (i).
\end{enumerate}
The exclusion of massive haloes has a strong impact on the halo clustering, thus it must be taken into account when the halo catalogues are constructed \citep{10.1046/j.1365-8711.2001.03894.x,2002MNRAS.333..730C,PhysRevD.88.083507,2015MNRAS.451.4266Z}. In our Universe, it is mainly caused by the facts that:
\begin{itemize}
    \item two or more haloes that are close enough can gravitationally collapse into a single more massive one;
    \item there is not enough matter to form multiple massive haloes on small scales.
\end{itemize}

For this method, the scaling factor $\gamma$ and the size of the grid $N_\mathrm{grid}$ are the two parameters that can be tuned to influence the halo and void clustering. Nevertheless, the effects of these parameters on the resulting void power spectrum are not straightforwardly interpretable.

\paragraph{CosmoGAME}
\label{sec:cosmogame}

Similarly to \textsc{SICKLE}, \textsc{CosmoGAME} estimates the density field in real space $\delta_\mathrm{m}(\mathbf{r})$ on which it assigns haloes. 
While $\delta_\mathrm{m}(\mathbf{r})$ is identical to the one estimated by \textsc{SICKLE} (except the $\gamma$ factor), the halo selection process and the tunable parameters are analogous to the galaxy assignment step for the Effective--Zel'dovich mocks \citep[\textsc{EZmocks}; ][]{10.1093/mnras/stu2301_ezmocks, 2021MNRAS.503.1149Z}. 
It is important to re-emphasize the fact that whilst \textsc{EZmocks} include the Zel'dovich approximation to estimate the gravitational evolution of the density field, \textsc{CosmoGAME} uses directly the Gaussian random field to assign haloes.

One of the \textsc{CosmoGAME}'s parameters used to select haloes is the critical density ($\delta_\mathrm{c}$). This variable plays the role of a threshold below which one cannot assign haloes \citep{2005A&A...443..819P} and thus has an impact on the three-point clustering of haloes \citep{10.1093/mnras/stv645}. 

After picking the density field values above $\delta_\mathrm{c}$, random numbers are added to them in order to take into account the stochasticity of the tracers \citep{10.1093/mnras/stu2301_ezmocks}:
\begin{equation}
    \delta_\mathrm{t}(\mathbf{r})=H(\delta_\mathrm{m} - \delta_\mathrm{c})\delta_\mathrm{m}(\mathbf{r})\times(1 + S),
\end{equation}
where:
\begin{equation}
\begin{split}
S=&\begin{cases}
G(\lambda), ~ G(\lambda) \geq 0;\\
\exp(G(\lambda)) - 1, ~ G(\lambda) < 0
\end{cases}
\end{split}
\end{equation}
and $H(x)$ is the Heaviside step function.
In the previous equation, $G(\lambda)$ is a random number sampled from a Gaussian distribution with a zero mean and a standard deviation $\lambda$ -- as a free parameter.

Lastly, a power-law probability density function (PDF) is used to assign haloes to the resulting density values:
\begin{equation}
    \mathcal{P}(n_\mathrm{t}) = A b^{n_\mathrm{t}},
\end{equation}
where $\mathcal{P}(n_\mathrm{t})$ is the probability to assign $n_\mathrm{t}$ haloes to a density peak. The fact that one has to ask for a fixed number of tracers puts a constrain on one of two parameters (i.e. $A$ or $b$). Thus, we fix $A$ (with $A>0$) and treat $b$ as the only free parameter within $0 < b < 1$. 
In practice, using the previous PDF, one computes the number of density values to which one should assign $n_\mathrm{t}$ tracers:
\begin{equation}
n_\mathrm{c}(n_\mathrm{t}) = \lfloor N_\mathrm{cell} \mathcal{P}(n_\mathrm{t}) \rceil,
\end{equation}
where $N_\mathrm{cell}=N_\mathrm{grid}^3$ ($N_\mathrm{grid}=512$, in this study) is the total number of cells in the density grid and the $\lfloor \cdot \rceil$ operator obtains the nearest integer.
Moreover, we compute the maximum number of haloes that can be possibly assigned to one density value as:
\begin{equation}
    n_\mathrm{t,~max} = \min_{n_\mathrm{t}>0}\{n_\mathrm{t}|N_\mathrm{cell} \mathcal{P}(n_\mathrm{t}) < 0.5\}.
\end{equation}

The tracer assignment is performed -- after the density values $\delta_\mathrm{t}(\mathbf{r})$ are sorted in descending order -- as follows:
\begin{enumerate}
    \item one assigns $n_\mathrm{t,~max}$ haloes to the highest $n_\mathrm{c}(n_\mathrm{t,~max})$ density values;
    \item one continues to assign $(n_\mathrm{t,~max} - i)$ haloes to the next $n_\mathrm{c}(n_\mathrm{t,~max} - i)$ density values, 
\end{enumerate}
where $i$ takes values from $1$ to $n_\mathrm{t,~max}$. The positions of the assigned haloes are sampled from a uniform distribution inside each of the grid cells. 

Another parameter of \textsc{CosmoGAME}, similarly to \textsc{SICKLE}, is the grid size $N_\mathrm{grid}$. Nonetheless, by adjusting the other parameters, one can emulate the effect of a different grid size. Thus, it is not used in the tuning process.

Lastly, \textsc{CosmoGAME} has been already run to create the void model power spectrum for the multi-tracer cosmological analysis with SDSS data by \citet{10.1093/mnras/stac390}.

\subsubsection{Parameter inference}
In order to infer the fitting parameters, we have written \textsc{pyBAOfit}\footnote{\url{https://github.com/Andrei-EPFL/pyBAOfit}}. The code uses a combination of \textsc{PyMultiNest}\footnote{\url{https://github.com/JohannesBuchner/PyMultiNest}} -- the \textsc{python} implementation of \textsc{MultiNest} \citep{2008MNRAS.384..449F, 2009MNRAS.398.1601F, 2019OJAp....2E..10F} -- and a Least-Square (LS) method \citep{Press2007, 10.1093/mnras/stac390} in order to decrease the computational time. While \textsc{PyMultiNest} samples the $(\alpha, B, \Sigma_\mathrm{nl}, c)$ parameters, the LS determines the best-fitting nuisance parameters $(a_0, a_1, a_2)$. \textsc{MultiNest} is a Bayesian Monte Carlo (MC) sampler, which provides not only the best-fitting parameters, but also the Bayesian evidence and the posterior distributions of the parameters. A more detailed discussion about the different treatment of the two sets of parameters is done in Section~\ref{sec:appendix_nuisance_parameters}.

The Bayesian inference is based on Bayes' theorem that provides a way to merge the prior information about the $\Theta$ parameters of a model $M$ with the measurements from the data $D$. Mathematically, the theorem provides the posterior probability density of the $\Theta$ parameters, given the data $D$ and the model $M$: 
\begin{equation}
    p(\Theta|D,\ M) = \frac{p(D|\Theta,\ M)p(\Theta|M)}{p(D|M)},
\end{equation}
where $p(\Theta|M)$ is the prior distribution of the $\Theta$ parameters (see Section~\ref{sec:param_priors}), $p(D|\Theta,\ M)$ is the likelihood -- related to the measurements from data $D$ -- and $p(D|M)$ is the Bayesian evidence -- $\mathcal{Z}$, a normalising factor and a valuable tool in model selection.

In the current study, we approximate the likelihood with a multivariate Gaussian:
\begin{equation}
    p(D|\Theta,\ M) = \mathcal{L}(\Theta)\sim\mathrm{e}^{-\chi^2(\Theta)/2},
\end{equation}
where $\chi^2$ is the chi-squared defined as:
\begin{equation}
    \chi^2(\Theta)=\textbf{v}^\mathrm{T}\textbf{C}^{-1}\textbf{v}.
\end{equation}
In the above formula, $\textbf{C}^{-1}$ is the inverse of the unbiased covariance matrix \citep{2007A&A...464..399H}, and $\textbf{v}$ is the difference between the model and the data vectors, i.e. $\textbf{v}=\boldsymbol{\xi}_\mathrm{data}-\boldsymbol{\xi}_\mathrm{model}(\Theta)$.

The unbiased covariance matrix $\textbf{C}$ is related to the sample covariance matrix of mocks $\textbf{C}_s$ as follows:
\begin{equation}
    \textbf{C}^{-1} = \textbf{C}_s^{-1}\frac{N_\mathrm{mocks} - N_\mathrm{bins} -2}{N_\mathrm{mocks} - 1}, 
\end{equation}
where $N_\mathrm{mocks}$ is the number of mocks used to compute the covariance matrix and $N_\mathrm{bins}$ is the length of the data vector $\boldsymbol{\xi}_\mathrm{data}$ included in the fitting process. Furthermore, $\textbf{C}_s$ can be decomposed into a multiplication between a matrix $\textbf{M}$ and its transpose:
\begin{equation}
    \textbf{C}_s = \frac{1}{N_\mathrm{mocks} - 1} \textbf{M}^\mathrm{T}\textbf{M}.
\end{equation}
Finally, the elements of the matrix \textbf{M} are computed as:
\begin{equation}
    \textbf{M}_{ij}=\xi_i(s_j) - \Bar{\xi}(s_j),~i = 1, 2, ..., N_\mathrm{mocks},~s_j\in[s_\mathrm{min}, s_\mathrm{max}], 
\end{equation}
where $\xi_i$ denotes the 2PCF of the $i-$th mock realisation, $\Bar{\xi}$ represents the mean 2PCF of all mocks and $[s_\mathrm{min}, s_\mathrm{max}]$ represents the interval of data points involved in the 2PCF fitting.

The quoted values of the parameters are the medians of the posterior distributions, and the $1\sigma$ uncertainties are half the differences between the 84th and 16th percentiles, unless otherwise specified.

\subsubsection{Parameter priors}
\label{sec:param_priors}
The Bayesian inference method requires prior knowledge about the measured parameters, generally implemented as a probability distribution function. In our case, we have mainly assumed uniform distributions $\mathcal{U}_{\left[a,\ b\right]}(\Theta)$:
\begin{equation}
\begin{split}
\mathcal{U}_{\left[a,\ b\right]}(\Theta)=&\begin{cases}
0, \hspace{2.65cm}   \Theta<a\\
\frac{1}{b-a}, \hspace{2.2cm}  \Theta\in[a, b]\\
0, \hspace{2.65cm}   \Theta>b.
\end{cases}
\end{split}
\end{equation}

While for the priors of $\alpha$ and $B$ we have generally imposed:
\begin{equation}
    p(\alpha)=\mathcal{U}_{\left[0.8,\ 1.2\right]}(\alpha),
\end{equation}
\begin{equation}
    p(B)=\mathcal{U}_{\left[0,\ 25\right]}(B),
\end{equation}
the prior of $\Sigma_\mathrm{nl}$ depends whether the 2PCF has been measured from boxes or from light-cones.
In the first case -- i.e. for boxes -- we implement a uniform prior:
\begin{equation}
    \label{eq:sigma_nl_prior}
    p\left(\Sigma_\mathrm{nl}\right)=\mathcal{U}_{\left[0,\ 30\right]\ \,h^{-1} \mathrm{Mpc}}(\Sigma_\mathrm{nl}).
\end{equation}
In the second case -- i.e. for light-cones --  we fix the values of $\Sigma_\mathrm{nl}$ to the ones in Table~\ref{tab:fixed_priors_light_cone}. The chosen intervals are large enough to not bias the measurements, as shown by \citep{ImproveBAOvoids_BOSS} and also obvious in Figures~\ref{fig:BAOfit_avg_16R_500_2pcf_cg}-\ref{fig:BAOfit_avg_16R_500_2pcf_xpar}.

\begin{table}
    \centering
    \begin{tabular}{c|c|c}
        \hline
             $\Sigma_\mathrm{nl} $ & Auto  & Cross \\
             $h^{-1} \mathrm{Mpc}$ &   &  \\
        \hline
            fix-$c$            & 9.03 & 9.77   \\
            PAR$_\mathrm{G}$   & 9.03 & 9.77   \\
            SK$_\mathrm{B}$    & 6.88 & 5.28   \\
            CG$_\mathrm{B}$    & 7.03 & 3.88   \\
            SK$_\mathrm{LC}$   & 7.68 & 6.77   \\
            CG$_\mathrm{LC}$   & 7.64 & 5.80   \\
        \hline
    \end{tabular}
    \caption{Prior values of $\Sigma_\mathrm{nl}$ when fitting the individual 2PCF of light-cones. These are the best-fitting values of the average of 1000 2PCF computed from light-cones. Cross -- void-galaxy cross 2PCF; Auto -- void auto 2PCF. \label{tab:fixed_priors_light_cone}}
\end{table}

The reason behind fixing the $\Sigma_\mathrm{nl}$ is that the light-cones have a smaller volume than the boxes, thus the corresponding 2PCF are noisier. Given the noisier 2PCF, $\Sigma_\mathrm{nl}$ is not properly constrained and the uncertainty of $\alpha$ is overestimated -- see also Figure~\ref{fig:A_alpha_avg_SK_realisations}. \citet{10.1093/mnras/stac390} have shown that fixing this parameter does not bias the measurements and thus it is appropriate to do it for the light-cones. 
In order to accurately measure $\Sigma_\mathrm{nl}$, we have fitted the average of all 1000 2PCF realisations measured from light-cones with a covariance matrix corresponding to the average 2PCF -- i.e. computed from 1000 realisations and rescaled by 1000 (rescaled covariance matrix) -- and the uniform prior shown in Eq.\,\eqref{eq:sigma_nl_prior}, as performed by \citet{10.1093/mnras/stac390}. The best-fitting $\Sigma_\mathrm{nl}$ values (Table~\ref{tab:fixed_priors_light_cone}) are then used in fitting the individual 2PCF from light-cones.

In the case of the parabolic model, as seen in Eq.\,\eqref{eq:parabolic_model}, there is an additional parameter $c$, for which we consider three cases:
\begin{itemize}
    \item a uniform prior for $c$  (PAR$_\mathrm{U}$)
\begin{equation}
    \label{eq:parabolic_uniform_prior}
    p(c)=\mathcal{U}_{\left[-10^4,\ 10^4\right]\ \,h^{-2} \mathrm{Mpc}^{2}}(c);
\end{equation}
    
    \item a uniform prior with two Gaussian tails (PAR$_\mathrm{G}$), similar to the one used in \citet{ImproveBAOvoids_BOSS}
\begin{equation}
\label{eq:par_g_prior_c}
\begin{split}
p(c) =&\begin{cases}
0, \hspace{2.65cm}   c<c_{\mathrm{min}}\\
A'\exp(-\frac{(c-c_{\mathrm{fmin}})^2}{2\sigma_c^2}), c\in[c_{\mathrm{min}}, c_{\mathrm{fmin}}]\\
A', \hspace{2.65cm}  c\in[c_{\mathrm{fmin}}, c_{\mathrm{fmax}}]\\
A'\exp(-\frac{(c-c_{\mathrm{fmax}})^2}{2\sigma_c^2}), c\in[c_{\mathrm{fmax}}, c_{\mathrm{max}}]\\
0, \hspace{2.65cm}   c>c_{\mathrm{max}},
\end{cases}
\end{split}
\end{equation}
where $c_{\mathrm{fmin}}=-100\,h^{-2} \mathrm{Mpc}^{2}$, $c_{\mathrm{fmax}}=900\,h^{-2} \mathrm{Mpc}^{2}$, $c_{\mathrm{min}}=-400\,h^{-2} \mathrm{Mpc}^{2}$, $c_{\mathrm{max}}=1200\,h^{-2} \mathrm{Mpc}^{2}$ and $\sigma_c=100\,h^{-2} \mathrm{Mpc}^{2}$;
    \item a fixed value of $c$ (fix $c$, see Table\,\ref{tab:fixed_c_priors}), as in \citep{10.1093/mnras/stac390}.
\end{itemize}
The uniform prior on $c$ (Eq.~\eqref{eq:parabolic_uniform_prior}) has been always used when we have fitted the average 2PCF (of 1000 realisations from LC and of 500 realisations from boxes). For the individual 2PCF, we have either fixed the values of $c$ -- as in Table~\ref{tab:fixed_c_priors} -- or used the PAR$_\mathrm{G}$ prior, Eq.~\eqref{eq:par_g_prior_c}.

\begin{table}
    \centering
    \begin{tabular}{c|c|c}
        \hline
            $c$ & Auto & Cross      \\
            $h^{-2} \mathrm{Mpc}^{2}$ &  &       \\
        \hline
            light-cone      & 2193    & 477 \\
            pre-recon box   & 1064    & 216 \\
            recon box       & 4030    & 319 \\
        \hline
    \end{tabular}
    \caption{Prior values of $c$ when fitting the individual 2PCF with a parabolic model. These are the best-fitting values of the average 2PCF (from 1000 light-cones or 500 boxes). Cross -- void-galaxy cross 2PCF; Auto -- void auto 2PCF. \label{tab:fixed_c_priors}}
\end{table}

Similarly to $\Sigma_\mathrm{nl}$,  we have determined the value of $c$ by fitting the average 2PCF (from 500 boxes or from 1000 light-cones) with the rescaled covariance matrix -- corresponding to the average 2PCF -- to mitigate the potential biases due to the cosmic variance of the mocks. The best-fitting values of $c$ -- shown in Table~\ref{tab:fixed_priors_light_cone} -- are used in the fitting of individual 2PCF. In contrast, to test the 2PCF fitting range, we use the covariance matrix corresponding to one 2PCF realisation (unscaled covariance matrix) together with the average 2PCF.

It is important to note that all the above priors have been used for fitting both the void auto-2PCF and the void-galaxy cross-2PCF. However, when we fit the void-galaxy cross-2PCF, we have to account for the negative bias of the DT voids \citep{10.1093/mnras/stw660}. Generally, the $B^2$ term in Eq.~\eqref{eq:ximodel} should be replaced by the product of the galaxy bias with the void one: $B_\mathrm{galaxy} \times B_\mathrm{void}$, with $B_\mathrm{void} < 0$. However, in this work, we do not write the explicit form because we do not fit simultaneously the void auto-2PCF, void-galaxy cross-2PCF and galaxy auto-2PCF.
Consequently, we simply replace $B^2$ with $-B^2$ in Eq.~\eqref{eq:ximodel} for the parabolic and the DW models. In contrast, the numerical models contain the information of the void negative bias in the shape of the resulting power spectrum, see the cross-clustering in Figure~\ref{fig:correct_template}.

\subsubsection{Model comparison}
In the next paragraphs, we define the parameters that we use to compare the models.

\paragraph{Bayes factor}
\label{sec:bayes_fac}
Apart from inferring parameters, Bayes' theorem can also be utilised to compare the quality of different models given prior probabilities of each models and their evidences:
\begin{equation}
    \frac{p(M_1|D)}{p(M_2|D)} = \frac{p(D|M_1)p(M_1)}{p(D|M_2)p(M_2)},
\end{equation}
where
\begin{equation}
    \mathcal{Z}_i \equiv p(D|M_i) = \int\mathcal{L}(\Theta)p(\Theta|M)\mathrm{d}\Theta
\end{equation}
is the Bayesian evidence, $p(M_1)/p(M_2)$ is the prior probability ratio between the two models and $p(M_1|D)/p(M_2|D)$ is the posterior probability ratio of the two models given the data set $D$.

\textsc{Multinest} provides the natural logarithm of the Bayesian evidence, thus one can easily compute $\ln\left(\mathcal{Z}_1/\mathcal{Z}_2\right)$, i.e. the natural logarithm of the Bayes factor between any two tested models. Given that we consider the prior probabilities of the models to be equal $p(M_1)=p(M_2)$, the Bayes factor is a direct indication of whether a model has a higher probability to be correct than another given a data set.

\paragraph{Tension parameter}
\label{sec:tension_param}
The most important aspect of a studied model is the capability to provide unbiased measurements of the Alcock--Paczynski parameter and its uncertainty. In order to have a quantitative description of the possible biases, we define the tension parameter $\tau(x, y| \sigma_x, \sigma_y)$ between two values $x$ and $y$, given their uncertainties $\sigma_x$ and $\sigma_y$, respectively:
\begin{equation}
    \label{eq:tension_parameter}
    \tau(x, y| \sigma_x, \sigma_y) = \frac{x - y}{\sqrt{\sigma_x^2 + \sigma_y^2}}.
\end{equation}
Naturally, this parameter can quantify the differences between different models, however it can also show the bias with respect to a reference. 

Given the fact that the input power spectrum of the \textsc{Patchy} mocks takes also the role of $P_\mathrm{lin}(k)$ in Eq.~\eqref{eq:templatevoids} to perform the BAO fitting, the expected measured value of $\alpha$ should be equal to one, in the absence of the non-linear evolution of the BAO peak and if all systematic effects are taken into account. Nonetheless, \citet{10.1093/mnras/stw312} has shown that the BAO can have a shift towards higher $\alpha$ values of $\sim0.25$ per cent for halo samples with linear bias from 1.2 to 2.8. Nevertheless, in this analysis, we approximate the reference to one and thus we also study the values of $\tau(\alpha, 1| \sigma_\alpha, 0)$.

\paragraph{Relative difference}
\label{sec:relative_difference}
We also formally define the relative difference in order to compare two quantities:
\begin{equation}
    \label{eq:relative_difference}
    \mathcal{R}(x, y) = 100 \times \left(\frac{x}{y} - 1\right).
\end{equation}
This tells us the difference in percentage between the two values.

\paragraph{Pull function}
In order to verify whether the uncertainties are correctly estimated, we define the pull function:
\begin{equation}
    \label{eq:pull_function}
    g(x) = \frac{x-\Bar{x}}{\sigma_x},
\end{equation}
where $\Bar{x}$ is the mean of a set of values $x$ and $\sigma_x$ is its standard deviation. 
If the histogram of the $g(x)$ values follow a standard normal distribution, one can conclude that the uncertainty of $x$ is correctly estimated.

%% file: files/results.tex
\section{Tests and Results}
\label{sec:results}

\subsection{Analysis and comparison of void clustering models}

\begin{figure}
 \includegraphics[width=\columnwidth]{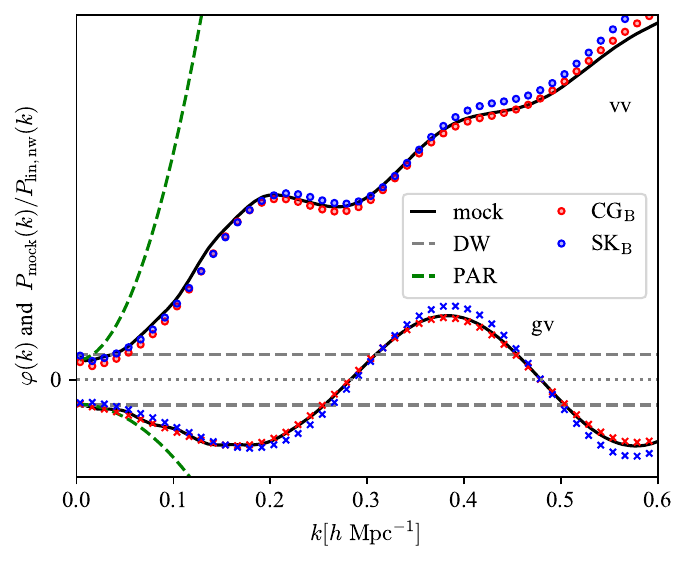}
 \caption{Comparison of $\varphi(k)$ -- defined in Eq.~\eqref{eq:additionalfactor} -- with the ratio between the average mock power spectrum $P_\mathrm{mock}$ and $P_\mathrm{lin,nw}$ (black). $\varphi(k)$ is computed for different models: grey dashed - de-wiggled model; green - parabolic model; red and blue - numerical models. $P_\mathrm{mock}$ is  obtained from 500 pre-reconstructed \textsc{Patchy} cubic mocks. The numerical models were re-scaled to match $P_\mathrm{mock}$, so the $y$ ticks are meaningless. See Table~\ref{tab:cg_sk_params} for the tuning parameters of the numerical models and Table~\ref{tab:model_abbreviations} for the abbreviations.}
 \label{fig:correct_template}
\end{figure}

We start by comparing the ratio $\varphi(k)$ Eq.~\eqref{eq:additionalfactor} of all models -- DW, PAR, \textsc{SICKLE}, \textsc{CosmoGAME} -- to the one of pre-reconstructed \textsc{Patchy} boxes. In Figure~\ref{fig:correct_template}, the colour dotted curves denote the numerical models, while the black curves represent the reference computed from 500 \textsc{Patchy} mocks. The horizontal dashed lines represent the DW model ($\varphi(k)=1$) that unequivocally under-fit the exclusion-effect-dominated reference. 
In contrast, one can observe that for small values of $k$ a parabola is a good approximation of the ratio, however it evidently fails for $k > 0.05 h~\mathrm{Mpc}^{-1}$. Unlike the previous models, the numerical models follow the reference up to $k=0.6\,h\mathrm{Mpc}^{-1}$. 

\begin{table}
    \centering
    \begin{tabular}{c|c|c|c|c|c}
        \hline
          & CG$_\mathrm{B}$ / CG$_\mathrm{LC}$ & CG$_\mathrm{def}$ & CG$_\mathrm{80}$ & CG$_\mathrm{120}$ \\
        \hline
            $\delta_c$ & 2.6  (1.8)  & 2.4  (1.6)  & 1.8  (1.2)  & 1.6  (1.8)  \\
            $\lambda$  & 1.0  (0.3)  & 2.0  (0.5)  & 0.4  (0.1)  & 1.5  (1.0)  \\
            $b$        & 0.44 (0.28) & 0.28 (0.20) & 0.32 (0.08) & 0.52 (0.28) \\
        \hline
                       & SK$_\mathrm{B}$ / SK$_\mathrm{LC}$  & SK$_\mathrm{def}$ &       &             \\
        \hline
            $N_\mathrm{grid}$     & 1024        & 1024        &             &             \\
            $\gamma$      & 0.075       & 0.3         &             &             \\
        \hline
    \end{tabular}
    \caption{Upper table: The values of the \textsc{CosmoGAME}'s free parameters used to create the numerical models. A more detailed description of the parameters can be found in Section~\ref{sec:cosmogame}. The abbreviations are defined in Table~\ref{tab:model_abbreviations}. The values in brackets are for the void-halo cross-power-spectrum, while the rest are for the void auto-power-spectrum. Lower table: The values of the \textsc{SICKLE}'s free parameters used to create the numerical models for both the void auto-power-spectrum and the void-halo cross-power-spectrum. More details can be found in Section~\ref{sec:sickle}}
    \label{tab:cg_sk_params}
\end{table}

\begin{figure}
 \includegraphics[width=\columnwidth]{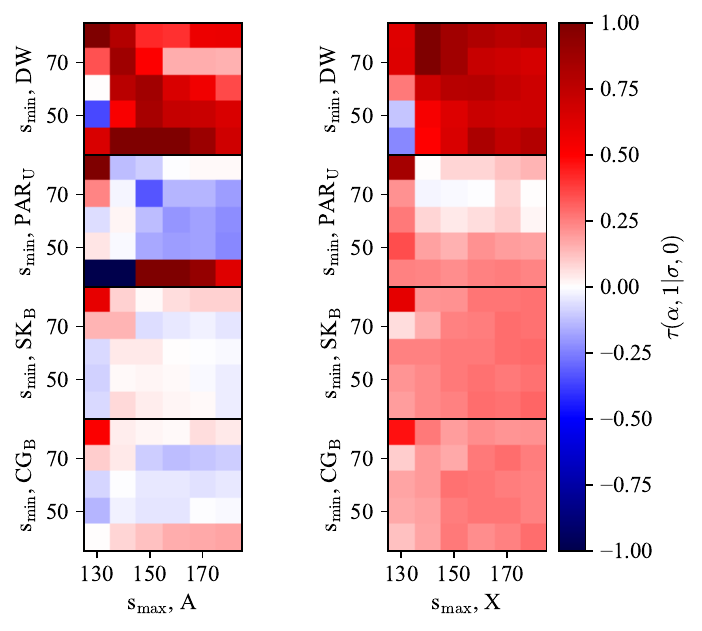}
 \caption{Comparison of different fitting ranges for four different models using $\tau(\alpha, 1| \sigma_\alpha, 0)$, Eq.~\eqref{eq:tension_parameter}. Both the average void auto-2PCF (left) and void-galaxy cross-2PCF (right) -- computed from 500 individual \textsc{Patchy} cubic mocks -- are considered. The abbreviations are defined in Table~\ref{tab:model_abbreviations}.}
 \label{fig:AX_B_CG_SK_GAL_PAR_fitting_interval}
\end{figure}

Furthermore, we check the robustness of all four models to the fitting range on the average correlation function -- computed from 500 mocks -- by evaluating the tension $\tau(\alpha, 1| \sigma_\alpha, 0)$. Figure~\ref{fig:AX_B_CG_SK_GAL_PAR_fitting_interval} contains the values of the tensions for the void auto-2PCF (left) and void-galaxy cross-2PCF (right) for different fitting intervals. 
Generally, the tension depends on the fitting range. However, its values are also influenced by the model and the studied clustering. 

Obviously, in the case of the de-wiggled model, the values of $\alpha$ are strongly biased for most fitting intervals, reaching values of $\sim1\sigma$ and above. This observation is consistent with the fact that this model is not suitable to describe the clustering of voids, as shown in \citet{ImproveBAOvoids_BOSS}.
The parabolic model shows significant improvements with respect to the de-wiggled model as most values are within $\pm0.2\sigma$ from zero. There are the clear outliers at $s_\mathrm{min}=40\,h^{-1}\mathrm{Mpc}$) for the void auto-2PCF, that do not appear for the void-galaxy cross-2PCF. An explanation might be that the exclusion effect in configuration space is present at smaller separations for the cross-clustering than for the auto-clustering.

The numerical models are more robust to the fitting ranges -- compared to the other methods -- given the fact that the tension of $\alpha$ is more homogeneous across the fitting ranges. There is the obvious exception of the narrow $s\in[80,130]\,h^{-1}\mathrm{Mpc}$ interval, which yields a strong bias given the lack of sufficient data points to describe well the peak. For most other fitting ranges, the results from the void auto-2PCF show little to no bias at all ($\pm0.1\sigma$), whilst a more consistent, yet not significant bias is present for the void-galaxy cross-2PCF ($\sim0.2 \sigma$). 

Due to the fact that around the $s\in[60, 150]\,h^{-1}\mathrm{Mpc}$ interval, the results are not sensitive to the fitting range, and this interval has been used in \citet{ImproveBAOvoids_BOSS}, we use it in the following tests. 

\begin{figure}
 \includegraphics[width=\columnwidth]{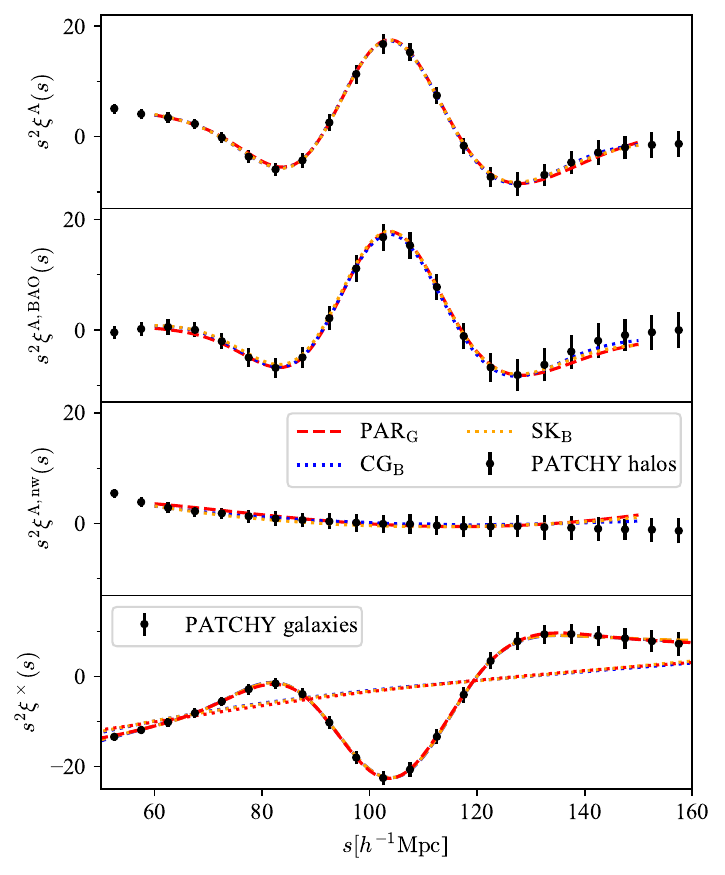}
 \caption{The best-fitting model curves for the average void auto-2PCF computed from 100 individual PATCHY halo boxes and for the average void-galaxy cross-2PCF computed from 500 individual \textsc{Patchy} galaxy boxes. First panel: the complete auto-2PCF. Second panel: the BAO peak (i.e. $s^2 \left[\xi(s) - \xi^\mathrm{nw}(s)\right]$). Third panel: the 2PCF without the BAO peak. The fourth panel: the complete cross-2PCF with the best-fitting curves (with and without BAO peak). The abbreviations are defined in Table~\ref{tab:model_abbreviations}.}
 \label{fig:AX_B_CG_SK_PAR_best_fit}
\end{figure}

Figure~\ref{fig:AX_B_CG_SK_PAR_best_fit} presents the best-fitting curves of the average correlation function for three models: parabolic model, SICKLE and CosmoGAME. All three models are describing well both the BAO peak and the broadband shape. Looking at the BAO-free best-fitting curves (the third panel and the dotted lines in the fourth panel of Figure~\ref{fig:AX_B_CG_SK_PAR_best_fit}), one can ascertain that none of the models introduce any additional signal at the position of the BAO peak.

\begin{figure}
 \includegraphics[width=\columnwidth]{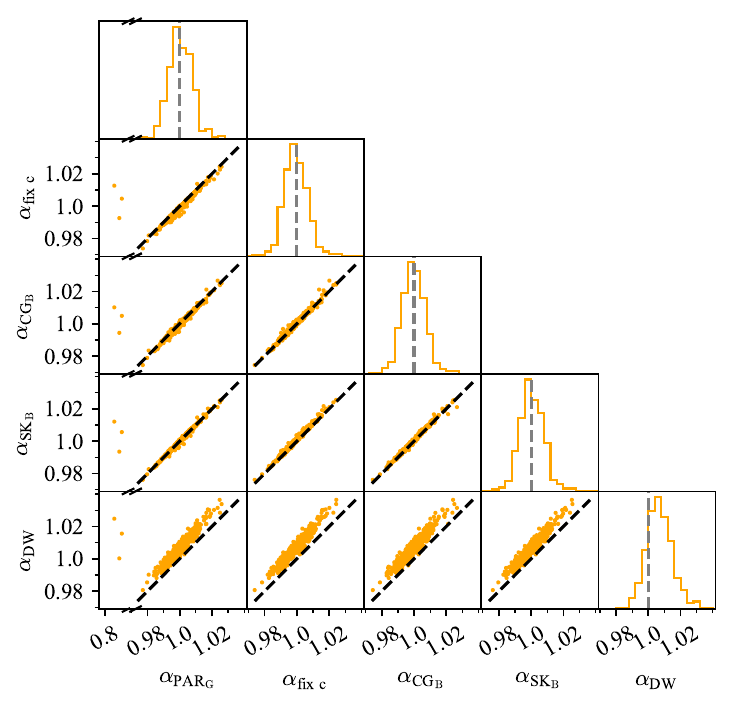}
 \caption{The $\alpha$ values obtained from the fitting of 500 individual void auto-2PCF computed from \textsc{Patchy} cubic mocks. The abbreviations are defined in Table~\ref{tab:model_abbreviations}. Grey - the theoretical value of 1; Black - the diagonal $y=x$.}
 \label{fig:A_B_CG_SK_GAL_PAR_fixc_corner_L}
\end{figure}

\begin{figure}
 \includegraphics[width=\columnwidth]{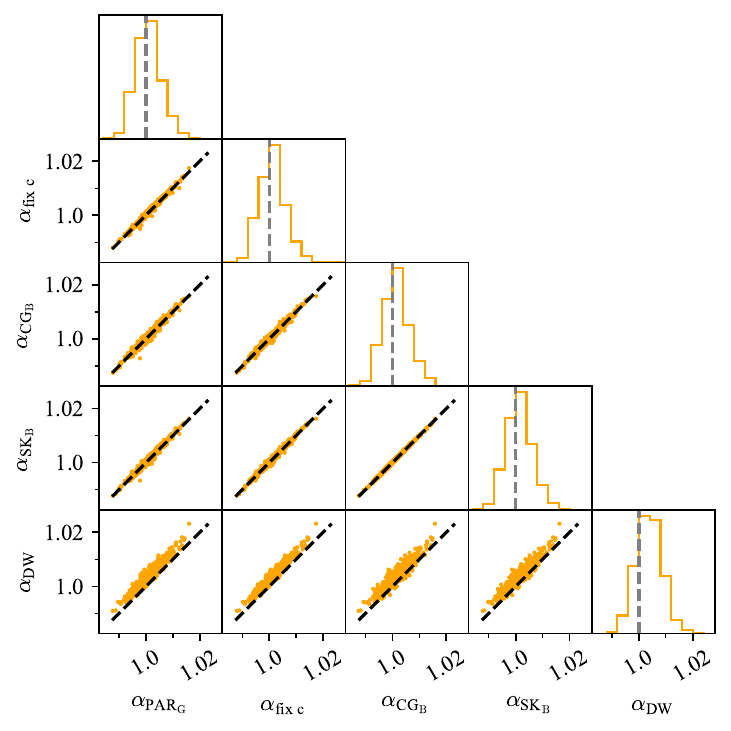}
 \caption{Same as Figure~\ref{fig:A_B_CG_SK_GAL_PAR_fixc_corner_L}, but for the void-galaxy cross-2PCF.}
 \label{X_B_CG_SK_GAL_PAR_fixc_corner_U}
\end{figure}

Figures~\ref{fig:A_B_CG_SK_GAL_PAR_fixc_corner_L} and \ref{X_B_CG_SK_GAL_PAR_fixc_corner_U} show a comparison of the four different models in terms of the measured $\alpha$ values from the 500 individual mocks. One can observe that the de-wiggled model induces a bias in the $\alpha$ values with respect to all other models for both void auto-2PCF and void-galaxy cross-2PCF.

The PAR$_\mathrm{G}$ model provides similar $\alpha$ values to the numerical models, but it is prone to fit poorly which leads to extreme values (the three points around the value of 0.8, in Figure~\ref{fig:A_B_CG_SK_GAL_PAR_fixc_corner_L}). In contrast, the parabolic model with fixed $c$ parameter is consistent with the numerical models for both the void auto-2PCF and the void-galaxy cross-2PCF. This suggests that a lack of a strong prior knowledge on $c$ presents risks of extreme failure. Consequently, we consider only the fixed-$c$ case in the further model comparison. Finally, the two numerical models are indistinguishable in terms of the resulting $\alpha$ values. 

\begin{figure}
 \includegraphics[width=\columnwidth]{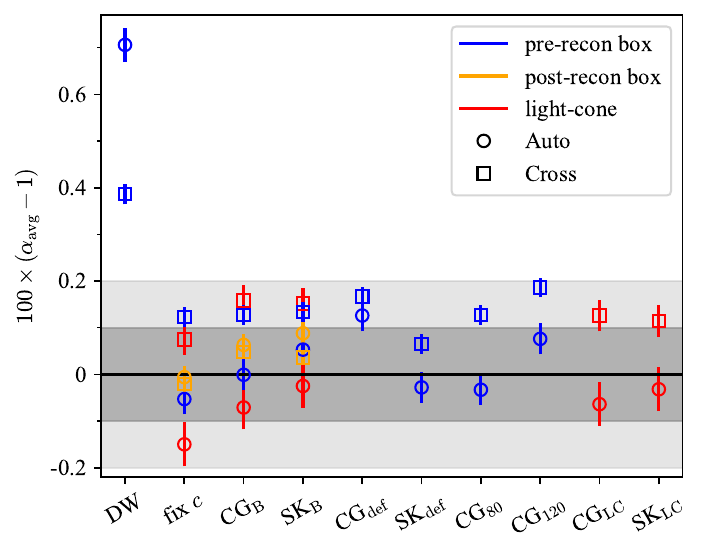}
 \caption{The average of 500 $\alpha$ values for \textsc{Patchy} boxes and of 1000 $\alpha$ values for \textsc{Patchy} light-cones  measured from void auto-2PCF and void-galaxy cross-2PCF. The error bars are computed as the standard deviation of the 500 (1000) $\alpha$ values further divided by $\sqrt{500}$ $(\sqrt{1000})$. The black horizontal denotes the values of zero, while the grey shaded areas encompass the intervals of $\pm0.2\%$ and $\pm0.1\%$ from the reference. See Table~\ref{tab:model_abbreviations} for abbreviations.}
 \label{fig:alpha_mean_std_mean_all}
\end{figure}

Analysing the average $\alpha$ of the 500 values from Figure~\ref{fig:alpha_mean_std_mean_all}, one can learn that the de-wiggled model introduces a bias of 0.4 to 0.7 per cent. In contrast, the bias shown by the numerical models and the parabolic model with the fixed $c$ is around $\pm0.1$ per cent for the void auto-2PCF and around 0.15 per cent for void-galaxy cross-2PCF. Moreover, the $\alpha$ values for the void auto-2PCF tend to be lower than one, while the values for the void-galaxy cross-2PCF larger than one. This is consistent with the findings of \citet{2013ApJ...763L..14M, 2018MNRAS.478.2495N}: due to the gravitational evolution, the clustering of over-dense regions underestimates the length of the sound horizon, whereas with the under-dense regions, the sound horizon is overestimated. Additionally, one has to consider that the values of $\alpha$ are slightly over-estimated, given the noise in the individual 2PCF and the large prior interval for $\Sigma_\mathrm{nl}$, as shown in Figure~\ref{fig:A_alpha_avg_SK_realisations}.

\begin{figure}
 \includegraphics[width=\columnwidth]{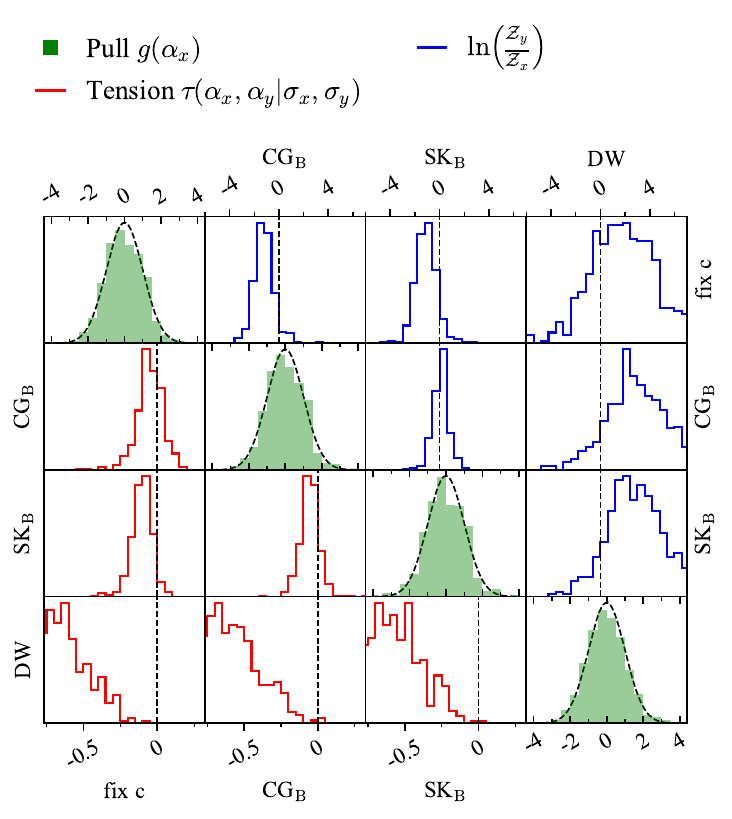}
 \caption{Diagonal panels: green - the histograms of the pull function $g(\alpha_x)$ values, Eq.~\eqref{eq:pull_function}; black - standard normal distributions. Lower triangular plots: the values of $\tau(\alpha_x, \alpha_y| \sigma_x, \sigma_y)$, Eq.~\eqref{eq:tension_parameter}, for all combinations of models. Upper triangular plot: the natural logarithm of the Bayes Factor $\ln{\left(\mathcal{Z}_y/\mathcal{Z}_x\right)}$ (see Section~\ref{sec:bayes_fac}). The results correspond to the individual fittings of the 500 void auto-2PCF computed from the \textsc{Patchy} cubic mocks. The abbreviations are defined in Table~\ref{tab:model_abbreviations}.} 
 \label{fig:AX_B_CG_SK_GAL_fixc_A_corner3_tension_evidence}
\end{figure}

\begin{figure}
 \includegraphics[width=\columnwidth]{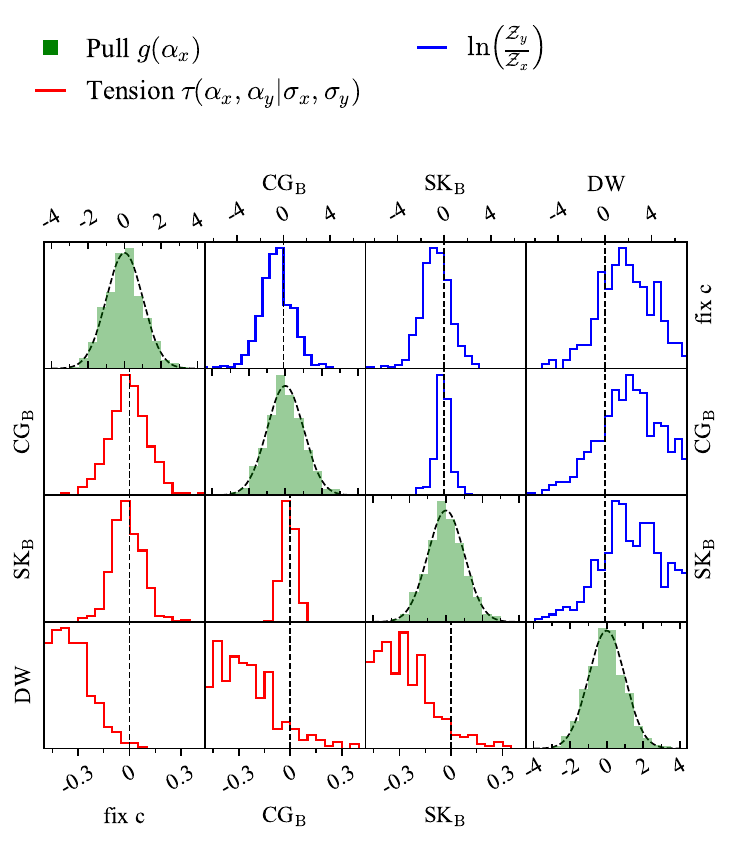}
 \caption{Same as Figure~\ref{fig:AX_B_CG_SK_GAL_fixc_A_corner3_tension_evidence}, but for the void-galaxy cross-2PCF.}
 \label{fig:AX_B_CG_SK_GAL_fixc_X_corner2_tension_evidence}
\end{figure}

In order to more robustly check the tensions between the models, we compute $\tau(\alpha_x, \alpha_y| \sigma_x, \sigma_y)$ between all pairs of models and show the resulting histograms in the lower triangular plots of Figures~\ref{fig:AX_B_CG_SK_GAL_fixc_A_corner3_tension_evidence} and \ref{fig:AX_B_CG_SK_GAL_fixc_X_corner2_tension_evidence}. The mean tensions with respect to the de-wiggled model reach values of $\sim-0.7\sigma$ for void auto-2PCF, and $\sim-0.5\sigma$ for void-galaxy cross-2PCF, supporting previous claims. Moreover, despite the important differences between the numerical models and the parabolic model with the fixed $c$ parameter observed in Figure~\ref{fig:correct_template}, the actual tensions between the measured $\alpha$ values are not significant (at most $\sim0.3 \sigma$ and on average $\sim0.1 \sigma$). This is because the damping term $a$ in the Hankel transform -- defined in Eq.~\eqref{eq:dampfouriertransform} -- decreases the amplitude of the models sharply at high $k$, and thus the higher $k$ discrepancies become less important.

While the tensions between the models can be informative on the possible introduced biases, the pull function $g(x)$ provides information about the uncertainty estimation. The resulting histograms can be observed along the diagonals of Figures~\ref{fig:AX_B_CG_SK_GAL_fixc_A_corner3_tension_evidence} and \ref{fig:AX_B_CG_SK_GAL_fixc_X_corner2_tension_evidence}. For both void auto-2PCF and void-galaxy cross-2PCF, one can estimate well the uncertainty $\sigma_\alpha$ with all models.

Finally, by studying the values of the Bayes factor for all pairs of models in the upper triangular panels of Figures~\ref{fig:AX_B_CG_SK_GAL_fixc_A_corner3_tension_evidence} and \ref{fig:AX_B_CG_SK_GAL_fixc_X_corner2_tension_evidence}, one can conclude that:
\begin{enumerate}
    \item the DW model is the least likely to be true;
    \item the parabolic model with a fixed $c$ is slightly disfavoured with respect to the numerical models;
    \item there is no preferential numerical model.
\end{enumerate}
These observations can be naturally interpreted by analysing Figure~\ref{fig:correct_template}:
\begin{enumerate}
    \item the DW model under-fits the exclusion wiggles;
    \item the parabolic model is a better description of the wiggles than DW, but worse than the numerical models;
    \item both numerical models follow similarly the exclusion feature up to $k=0.6\,h\mathrm{Mpc}^{-1}$.
\end{enumerate}

\subsection{Robustness tests against systematic errors}
\label{sec:box_systematic_error}

\begin{figure}
 \includegraphics[width=\columnwidth]{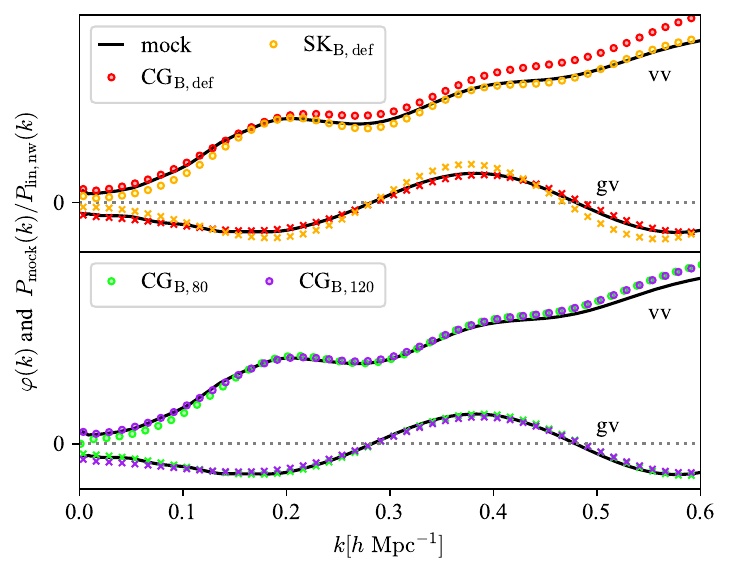}
 \caption{Same as Figure~\ref{fig:correct_template}, but with different models. Upper panel: defectively calibrated numerical models. Lower panel: calibrated numerical models that are obtained from halo catalogues with a number density of $80\%$ and $120\%$ of the reference number density.}
 \label{fig:wrong_imp_80_120_template}
\end{figure}

In this section, we investigate the sensitivity of BAO measurements to possible systematic errors in the numerical models and the data. 
Initially, we examine the sensitivity of the measured $\alpha$ to the parameters of \textsc{CosmoGAME} and \textsc{SICKLE} by shifting them away from the fiducial values (see Table~\ref{tab:cg_sk_params}). As a result, the newly computed power spectra (defective models, SK$_\mathrm{def}$, CG$_\mathrm{def}$, see Figure~\ref{fig:wrong_imp_80_120_template}) do not describe as well as the fiducial ones the reference clustering.

The second set of tests evaluates the robustness of the numerical models to potentially uncorrected systematic effects in the data. For example, the galaxy number density along the redshift is assumed to be isotropic, however, there are inhomogeneities across that sky, which means that the local number density of galaxies is not everywhere correctly estimated \citep[see e.g. Appendix A of ][]{2021MNRAS.503.1149Z}.
This is important because a different matter density yields a different void size distribution \citep{10.1093/mnras/stw660, 2022MNRAS.513.5407F} that finally alters the exclusion pattern \citep{10.1093/mnras/stw884}.

Another example of a systematic effect is the incompleteness in the data-set. For the SDSS data, on average, the incompleteness is lower than $5$ per cent. In some sectors, the incompleteness can get as large as $50$ per cent, but those regions cover small areas \citep{2016MNRAS.455.1553R,10.1093/mnras/staa2416}. Normally, these effects are included in the random and mock catalogues so that they compensate the ones in the data. However, the estimation of the galaxy number density might be imprecise, so the incompleteness effect might not be entirely removed. Consequently, we emulate these imprecise estimations by re-calibrating both codes’ parameters (see
Table~\ref{tab:cg_sk_params}), while asking for a halo number density that is different than the reference by $-20$ per cent (CG$_{80}$) and +20 per cent (CG$_{120}$).
These considered differences are fairly conservative compared to the expected errors in galaxy density estimations.

\begin{figure}
 \includegraphics[width=\columnwidth]{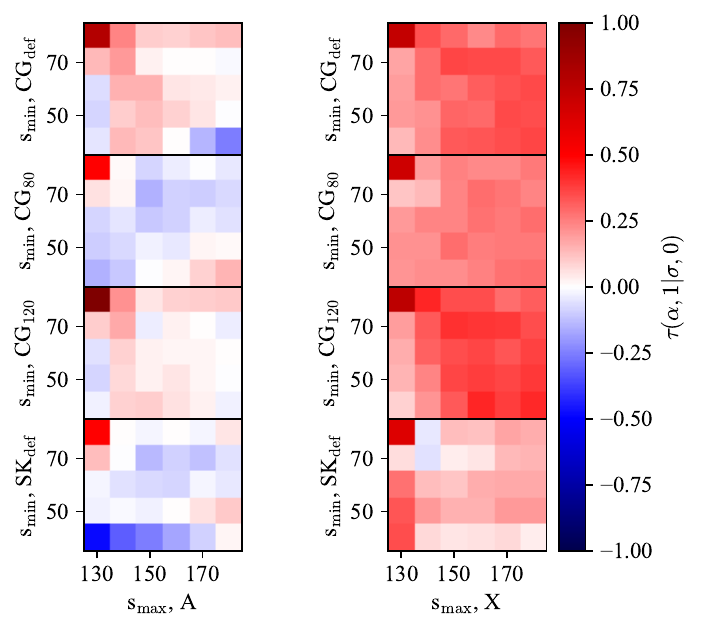}
 \caption{Comparison of different fitting ranges for four different cases using $\tau(\alpha, 1| \sigma, 0)$. Both the average void auto-2PCF (left) and void-galaxy cross-2PCF (right) -- computed from 500 individual \textsc{Patchy} cubic mocks -- are considered. The abbreviations are defined in Table~\ref{tab:model_abbreviations}.}
 \label{fig:AX_B_CGimp_GC80_GC120_SKimp_fitting_interval}
\end{figure}

Figure~\ref{fig:AX_B_CGimp_GC80_GC120_SKimp_fitting_interval} shows how the numerical models shown in Figure~\ref{fig:wrong_imp_80_120_template} perform when the average void auto-2PCF (left) and the average void-galaxy cross-2PCF (right) from 500 mocks are fitted in different fitting ranges.
On one hand, for the void auto-2PCF, the defective numerical models have generally a slightly larger bias compared to the fiducial ones (Figure~\ref{fig:AX_B_CG_SK_GAL_PAR_fitting_interval}), however most values remain within $\pm0.1\sigma$ from zero. On the other hand, for the void-galaxy cross-2PCF, CG$_\mathrm{def}$ imposes a stronger bias on the measurement of $\alpha$ ($\sim 0.35 \sigma$) than CG$_\mathrm{B}$, whereas SK$_\mathrm{def}$ decreases the bias from $\sim0.2\sigma$ (SK$_\mathrm{B}$) to $\sim 0.15\sigma$.
In the case of the void auto-2PCF, CG$_\mathrm{80}$ and CG$_\mathrm{120}$ remain within $\pm0.1\sigma$ bias from zero. For the void-galaxy cross-2PCF, the bias induced by CG$_\mathrm{80}$ is similar to the fiducial case, while CG$_\mathrm{120}$ increases the bias to $\sim0.3\sigma$.

\begin{figure}
 \includegraphics[width=\columnwidth]{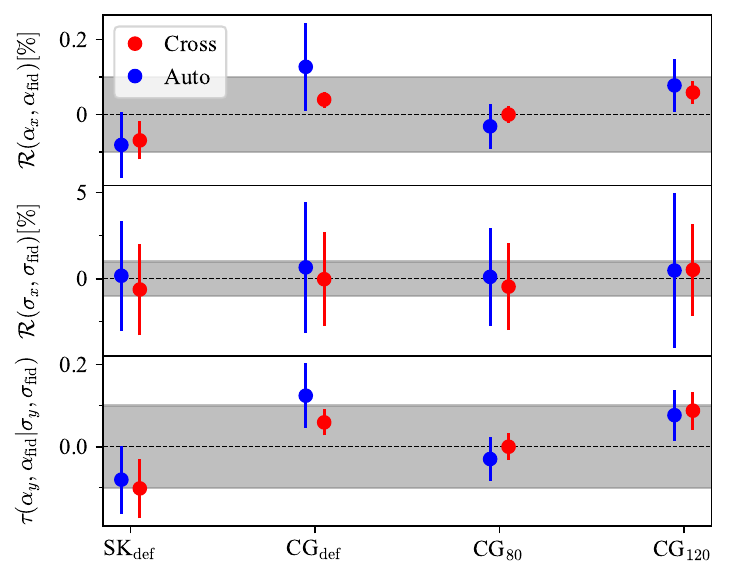}
 \caption{Comparison between the error-affected numerical models and the corresponding fiducial ones --  CG$_\mathrm{B}$ or SK$_\mathrm{B}$ --   using the clustering (blue for void auto-2PCF; red for void-galaxy cross-2PCF) computed from 500 individual pre-reconstructed \textsc{Patchy} cubic mocks. The abbreviations are defined in Table~\ref{tab:model_abbreviations}. First two panels: the relative difference (Eq.~\eqref{eq:relative_difference}) between the $\alpha$ values and $\sigma_\alpha$ values, respectively. Last panel: the tension from Eq.~\eqref{eq:tension_parameter}. The shown values and error bars are the averages and the standard deviations of 500 individual measurements. From top to bottom, the shaded areas delineate $\pm0.1\%$, $\pm1\%$ and $\pm0.1\sigma$, respectively.}
 \label{fig:AX_B_CG_CG80_CGimp_CG120_SK_SK_imp}
\end{figure}

Figure~\ref{fig:AX_B_CG_CG80_CGimp_CG120_SK_SK_imp} contains a comparison between the results of the fiducial CG$_\mathrm{B}$ model and the CG$_\mathrm{def}$, CG$_\mathrm{80}$ and CG$_\mathrm{120}$ ones, for void auto-2PCF (in blue) and void-galaxy cross-2PCF (in red). In the case of the void auto-2PCF, the strongest tension occurs between CG$_\mathrm{B}$ and CG$_\mathrm{def}$, i.e. $\sim0.15$ per cent or $\sim0.15\sigma$ on average. In terms of the $\sigma_\alpha$ values, these three models are consistent with the fiducial CG$_\mathrm{B}$ within $\pm1$ per cent on average.

For \textsc{SICKLE}, we have only tested the sensitivity to the tuning parameters and we present the results in Figure~\ref{fig:AX_B_CG_CG80_CGimp_CG120_SK_SK_imp}. The bias introduced by SK$_\mathrm{def}$ with respect to the fiducial SK$_\mathrm{B}$ is on average $-0.1$ per cent or $-0.1\sigma$. In contrast, the uncertainties are consistent with fiducial case within $\pm1$ per cent on average, as for CG.

Analysing the results of CG$_\mathrm{def}$, CG$_\mathrm{120}$, CG$_\mathrm{80}$ and SK$_\mathrm{def}$ in Figure~\ref{fig:alpha_mean_std_mean_all}, the average of the 500 $\alpha$ values is within $\sim\pm0.1$ per cent from the reference for four cases, while for the other four cases the bias is lower than $\sim0.2$ per cent. This suggests that even for larger survey such as DESI, the numerical models are robust enough to provide unbiased measurements of $\alpha$.

\subsection{Robustness tests against BAO reconstruction}
\begin{figure}
 \includegraphics[width=\columnwidth]{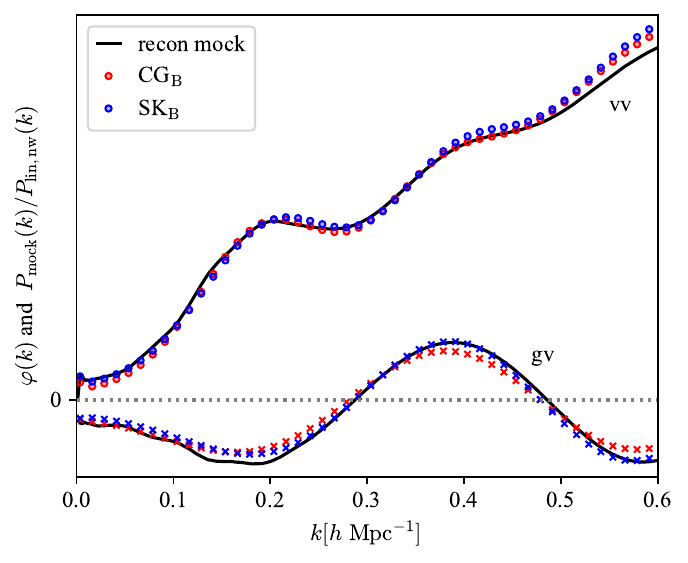}
  \caption{Same as Figure~\ref{fig:correct_template}, but a different $P_\mathrm{mock}$ and only CG$_\mathrm{B}$ and SK$_\mathrm{B}$. $P_\mathrm{mock}$ is computed from 500 \textsc{Patchy} reconstructed cubic mocks.}
 \label{fig:correct_template_post_recon_data}
\end{figure}

Figure~\ref{fig:correct_template_post_recon_data} shows a comparison between the average power spectrum of 500 reconstructed \textsc{Patchy} catalogues and the numerical models presented in Figure~\ref{fig:correct_template}, for both void auto-2PCF and void-galaxy cross-2PCF. It suggests that CG$_\mathrm{B}$ and SK$_\mathrm{B}$ can describe well the void clustering and be employed in BAO analysis.

\begin{figure}
 \includegraphics[width=\columnwidth]{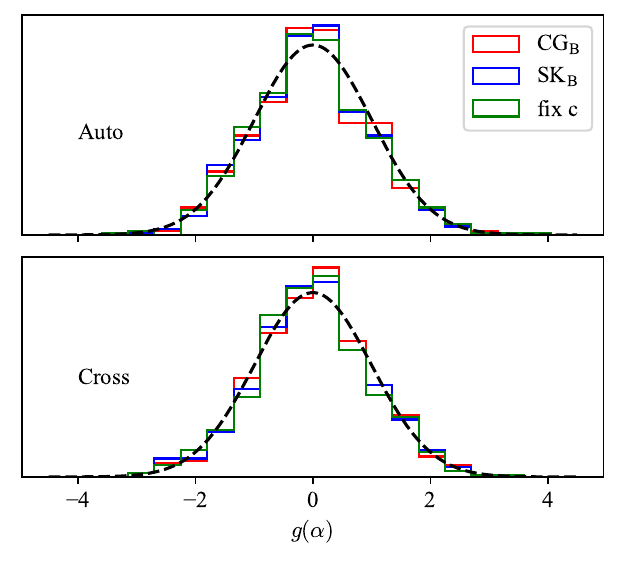}
 \caption{The histogram of the 500 pull $g(\alpha)$ values obtained from the individual fittings of the void auto-2PCF (upper panel) and void-galaxy cross-2PCF (lower panel) computed from reconstructed \textsc{Patchy} cubic mocks. The results are obtained using the two numerical models and the parabolic model with a fixed $c$ parameter (coloured histograms, see Table~\ref{tab:model_abbreviations} for abbreviations.). The black dashed line represents a standard normal distribution.}
 \label{fig:A_B_CGrec_SKrec_tensionone}
\end{figure}

After fitting the 500 individual void auto-2PCF (upper panel) and 500 void-galaxy cross-2PCF (lower panel), we compute the histogram of the pull $g(\alpha)$ values shown in Figure~\ref{fig:A_B_CGrec_SKrec_tensionone}. In both cases, the distributions are consistent with a standard normal one (black dashed line), meaning fix $c$, CG$_\mathrm{B}$ and SK$_\mathrm{B}$ provide correct estimations of $\sigma_\alpha$. Moreover, looking at Figure~\ref{fig:alpha_mean_std_mean_all}, the $\alpha_\mathrm{avg}$ values corresponding to three previous models (orange points) are within $\pm0.1$ per cent from the reference. One can also notice that for the void-galaxy cross-2PCF, the bias has systematically decreased by applying reconstruction on the galaxy catalogues, strengthening the observations of \citet{2013ApJ...763L..14M, 2018MNRAS.478.2495N} that the gravitational evolution shifts the BAO peak of galaxies to lower separation.

Considering the fact that the reconstruction inverts the effect of the gravitational evolution and that the numerical models are based on Gaussian random fields -- without any gravitational evolution -- these models should describe better the reconstructed data. Thus, one should ideally calibrate the \textsc{CosmoGAME} and \textsc{SICKLE} for both post and pre-reconstructed data. Nonetheless, the current results show that the same set of void model power spectra (CG$_\mathrm{B}$ and SK$_\mathrm{B}$) can be used in both scenarios.

\subsection{Robustness tests against survey-geometry effects}
\label{sec:survey-geometry_effects}

\begin{figure}
 \includegraphics[width=\columnwidth]{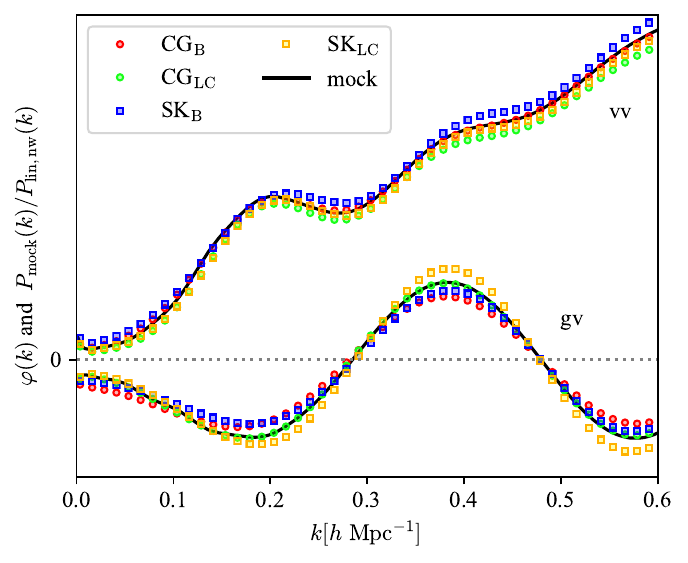}
  \caption{Same as Figure~\ref{fig:correct_template}, but a different $P_\mathrm{mock}$ and additionally CG$_\mathrm{LC}$ and SK$_\mathrm{LC}$. $P_\mathrm{mock}$ is computed from 1000 \textsc{Patchy} light-cone mocks}
 \label{fig:LC_template}
\end{figure}

In this subsection, we investigate the performance and robustness of the numerical models on light-cone data (described in Section~\ref{sec:data_lc}). Given the smaller volume of the light-cone compared to the box, the correlation functions are noisier. Consequently, we have used 1000 \textsc{Patchy} realisations to reduce the noise.
We have created two additional numerical models ($\mathrm{CG}_\mathrm{LC}$ and $\mathrm{SK}_\mathrm{LC}$) by applying the survey-geometry on the cubic catalogues corresponding to $\mathrm{CG}_\mathrm{B}$ and $\mathrm{SK}_\mathrm{B}$. The resulting void model power spectra are shown in Figure~\ref{fig:LC_template}.

\begin{figure}
 \includegraphics[width=\columnwidth]{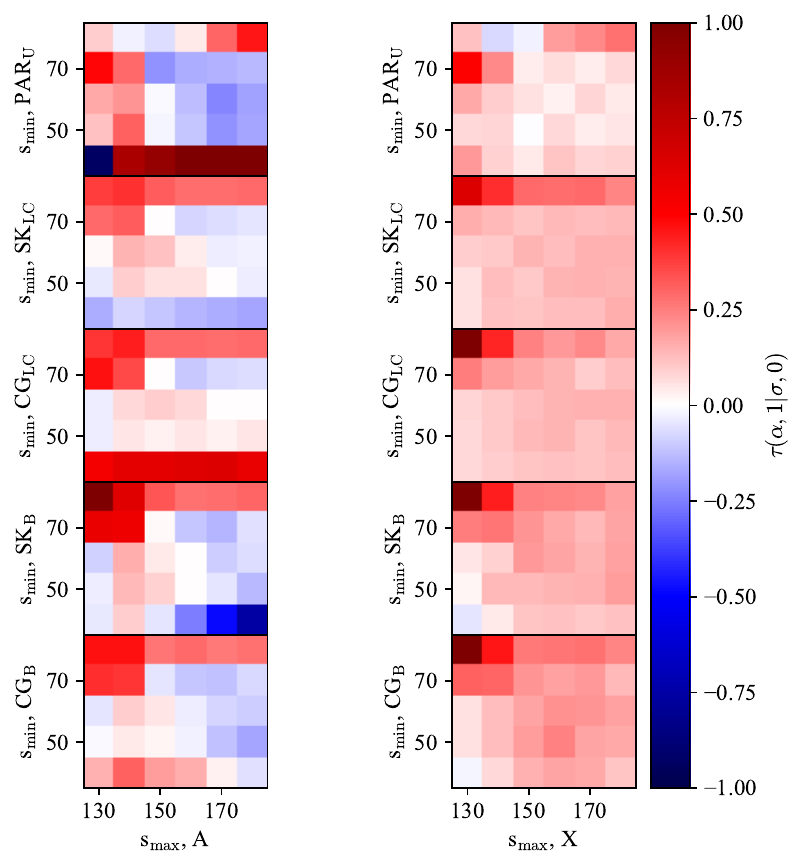}
 \caption{Comparison of different fitting ranges for five different cases using $\tau(\alpha, 1| \sigma, 0)$. Both the average void auto-2PCF (left) and void-galaxy cross-2PCF (right) -- computed from 1000 individual \textsc{Patchy} light-cone mocks -- are considered. The abbreviations are defined in Table~\ref{tab:model_abbreviations}.}
 \label{fig:LC_fitting_ranges}
\end{figure}

At this stage, we only test $\mathrm{CG}_\mathrm{B}$, $\mathrm{SK}_\mathrm{B}$, $\mathrm{CG}_\mathrm{LC}$, $\mathrm{SK}_\mathrm{LC}$ and the parabolic model, given that the DW model is obviously insufficient to describe voids. Figure~\ref{fig:LC_fitting_ranges} shows similar results as Figure~\ref{fig:AX_B_CG_SK_GAL_PAR_fitting_interval}, most biases for the void auto-2PCF are within $[-0.1, 0.1]\sigma$ interval, while for the void-galaxy cross-2PCF, most values of the tension are lower than $+0.2\sigma$.

Studying the average of the 1000 $\alpha$ values in Figure~\ref{fig:alpha_mean_std_mean_all}, we observe that five points -- $\mathrm{SK}_\mathrm{LC}$, $\mathrm{CG}_\mathrm{LC}$, $\mathrm{CG}_\mathrm{B}$, $\mathrm{SK}_\mathrm{B}$ for auto-2PCF and fix $c$ for cross-2PCF)-- are within $\pm0.1$ per cent from the reference, while the remaining five are within $\pm0.2$ per cent.

\begin{figure}
 \includegraphics[width=\columnwidth]{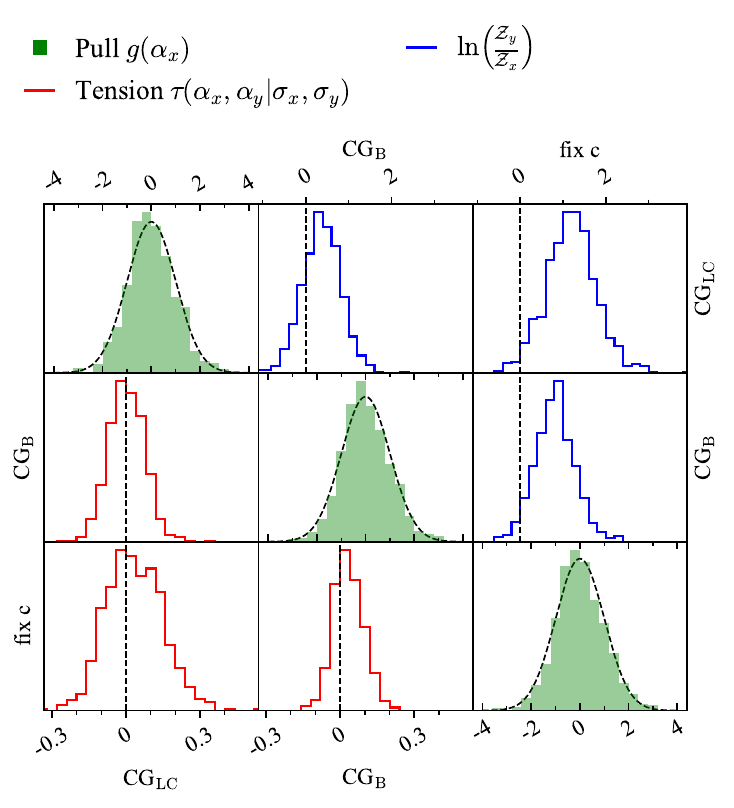}
 \caption{Same as Figure~\ref{fig:AX_B_CG_SK_GAL_fixc_A_corner3_tension_evidence}, but for 1000 void auto-2PCF computed from the \textsc{Patchy} light-cone mocks.}
 \label{fig:LC_evi_tension_pullone_auto}
\end{figure}

\begin{figure}
 \includegraphics[width=\columnwidth]{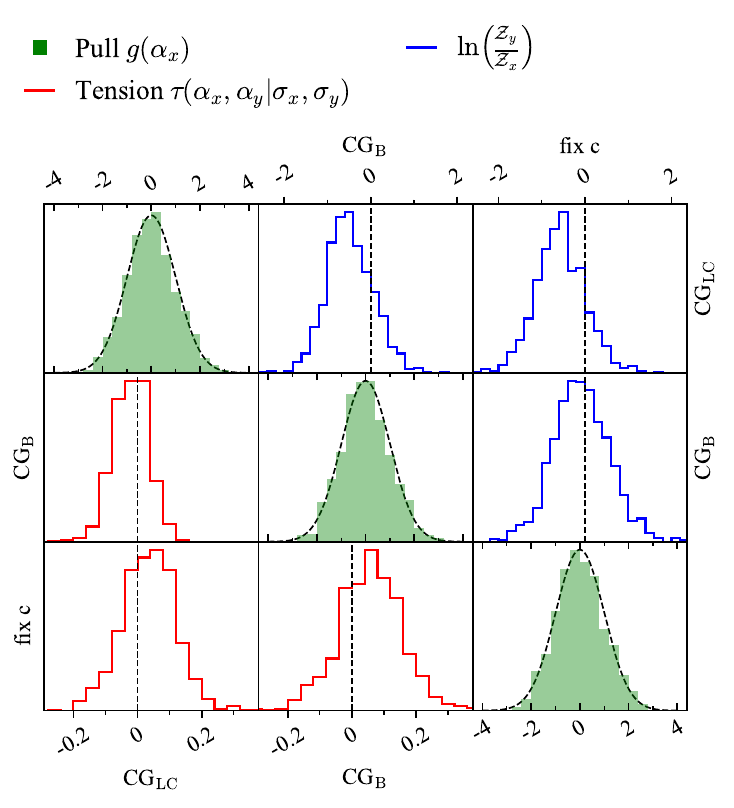}
 \caption{Same as Figure~\ref{fig:AX_B_CG_SK_GAL_fixc_A_corner3_tension_evidence}, but for 1000 void-galaxy cross-2PCF computed from the \textsc{Patchy} light-cone mocks.}
 \label{fig:LC_evi_tension_pullone_cross}
\end{figure}

Analysing the tension parameter between the $\mathrm{CG}_\mathrm{LC}$, $\mathrm{CG}_\mathrm{B}$ and the fix-$c$ models in Figures~\ref{fig:LC_evi_tension_pullone_auto}, \ref{fig:LC_evi_tension_pullone_cross}, we observe that there is no significant tension:  the mean values of the histograms are at most $0.1 \sigma$ from 0, while the highest deviations are $\sim0.3\sigma$. Moreover, the histograms of the 1000 pull $g(\alpha)$ values -- diagonal panels of the same figures -- additionally show that the uncertainties of $\alpha$ are correctly estimated by all models.

In terms of the most probable model for the void auto-2PCF, the logarithm of the Bayes Factor -- upper diagonal panels of Figure~\ref{fig:LC_evi_tension_pullone_auto} -- suggests that the parabolic model with a fixed $c$ parameter is slightly disfavoured against the numerical models. Furthermore, the light-cone numerical model is slightly preferred compared to the one constructed for boxes. In contrast, the results from void-galaxy cross-2PCF -- Figure~\ref{fig:LC_evi_tension_pullone_cross} -- show that the parabolic model is slightly favoured with respect to the numerical models. Moreover, it shows that $\mathrm{CG}_\mathrm{LC}$ is slightly disfavoured against the $\mathrm{CG}_\mathrm{B}$.

We only show the results of \textsc{CosmoGAME} due to visibility reasons, however we have also analysed the results of \text{SICKLE} in Appendix~\ref{sec:light-cone_results} and shown that the same conclusions are available in this case. Moreover, there is no preference between the CG$_\mathrm{LC}$ and SK$_\mathrm{LC}$, nor between CG$_\mathrm{B}$ and SK$_\mathrm{B}$.

%% file: files/conclusion.tex
\section{Conclusion}
We have introduced two numerical techniques to model the DT void clustering: \textsc{CosmoGAME} and \textsc{SICKLE}. The main steps to construct the models are the following:
\begin{itemize}
    \item the initial conditions are built starting from a BAO free linear power spectrum;
    \item haloes are assigned directly on the density field corresponding to the initial conditions;
    \item voids are detected using \textsc{DIVE};
    \item the void power spectrum is computed.
\end{itemize}
The difference between the two techniques lays into to the halo assignment process on the density field.

Furthermore, we have compared the performance of the two numerical models with the de-wiggled model of galaxies and a parabolic model introduced by \citet{ImproveBAOvoids_BOSS} for the BAO analysis with DT voids. To this end, we have used 500 \textsc{Patchy} cubic mocks and 1000 \textsc{Patchy} light-cone mocks \citep[similar to the BOSS DR12 LRG sample; ][]{2015ApJS..219...12A}. On one hand, the de-wiggled model can bias the measurements of $\alpha$ by $0.4$ to $0.7$ per cent on average, when fitting the 2PCF from boxes. Thus, as also shown in \citet{ImproveBAOvoids_BOSS}, the de-wiggled model is not a viable model for voids. On the other hand, the parabolic model can provide unbiased results, however it tends to provide outlier values of $\alpha$ when the additional parameter $c$ is not fixed. As a result, one has to fit the average of multiple mock 2PCF to precisely measure the value of $c$, so that it can be fixed when fitting individual 2PCF. Given that the cosmology of the mocks can be different from the one of the measured data, this might introduce a bias when fitting the clustering of data. In contrast, the numerical models can be directly calibrated on the void power spectrum computed from the measured data, as the exclusion pattern is much stronger than the noise.

By fitting the individual 2PCF from boxes, we have observed that the numerical models and the fixed $c$ parabolic model are in agreement within $\sim0.1\sigma$. Moreover, the histograms of the 500 values of $g(\alpha)$ are consistent with a standard normal distribution, meaning that all models estimate correctly the uncertainty of $\alpha$. For the void auto-2PCF, the three models provide $\alpha$ values within $\pm0.1$ per cent from the reference, while for void-galaxy cross-2PCF the bias is below $\sim 0.15$ per cent. Studying the Bayes factor, the two numerical methods are favoured with respect to the parabolic model and there is no preferred numerical technique. Finally, the results provided by the two new models are less affected by the fitting range than the parabolic model.

We have analysed the robustness of the two numerical techniques to systematic errors such as incompleteness and defective calibration. The average of the 500 $\alpha$ values is within $\sim 0.2$ per cent from the reference value for all four cases affected by systematic effects. Thus, we can conclude that \textsc{CosmoGAME} and \textsc{SICKLE} are resilient to such systematic errors.

Given the fact that the BAO reconstruction is a standard procedure in BAO analysis, we study the behaviour of the two newly introduced techniques and the fixed $c$ parabolic model on the reconstructed \textsc{Patchy} catalogues. We have observed that the values of $\alpha$ are consistent with one within $\pm0.1$ per cent and the uncertainty is well estimated, implying that \textsc{CosmoGAME} and \textsc{SICKLE} can be employed in modelling voids from both reconstructed and pre-reconstructed data-sets.

Lastly, we have tested \textsc{CosmoGAME}, \textsc{SICKLE} and the fixed $c$ parabolic model on light-cones. In this case, the numerical models based on boxes have similar performances as the ones based on light-cones, i.e. uncertainties are well estimated and no tension between the models have been noticed. Slight discrepancies occur between the void auto-2PCF and void-galaxy cross-2PCF cases in terms of Bayes factors. For the void-auto 2PCF, the light-cone based numerical models have a higher evidence than the box based ones and all void model power spectra are more likely to be correct than the parabolic model with a fixed $c$. In contrast, for void-galaxy cross-2PCF, the numerical models based on light-cones are slightly disfavoured against the ones based on boxes and the parabolic model. Analysing, the average of 1000 $\alpha$ values, we have noticed that most void model power spectra provide results within $\pm0.1$ per cent from the reference and all of them are within $\pm0.2$ per cent. This suggests that there is no bias introduced by the numerical models.

Even though, in the current case, the parabolic model with fixed $c$ parameter has similar performances to the numerical models -- in terms of estimating the $\alpha$ and its uncertainty -- \citet{2022arXiv220806238T} have explained that for void quasars, that have a much stronger exclusion at even larger scales, the parabolic model cannot be used anymore. Therefore a better description of the void exclusion is necessary and the two numerical models can provide it. Moreover, the numerical models have the potential for even smaller biases due to the possibility of fine tuning the parameters to reach a better agreement at large values of $k$. 

Finally, as explained by \citet{ImproveBAOvoids_BOSS, 10.1093/mnras/stac390}, the combined 2PCF of voids and galaxies is preferred over multiple 2PCF due to a lower dimension of the data vector and thus a smaller required number of mocks. Consequently, for future studies, we will adapt the numerical models to the combined 2PCF for a multi-tracer cosmological analysis. 

In conclusion, the usage of \textsc{CosmoGAME} or \textsc{SICKLE} in a BAO analysis with DT voids provides robust and unbiased measurements of the Alcock-Paczynski parameter. Moreover, the Bayes factor indicates a higher probability of these models to be true compared to the parabolic one. Nevertheless, we foresee the utility of these numerical methods in the study of different kind of voids or for different properties: e.g. void density contrast.

%% file: files/appendix.tex
\section{Reducing the noise of the numerical models}
\label{sec:templatecreation}

Given the fact that each halo and void catalogues produced by \textsc{CosmoGAME} and \textsc{SICKLE} has an intrinsic noise, the measured power spectrum and its Hankel transform Eq.~\eqref{eq:dampfouriertransform} are not smooth. In this section, we analyse how the number of realisations used to compute the void model power spectrum ($P_\mathrm{t, nw}(k)$) and the value of the damping factor $a$ affect the Hankel transform of $P_\mathrm{t, nw}(k)$.

\begin{figure}
 \includegraphics[width=\columnwidth]{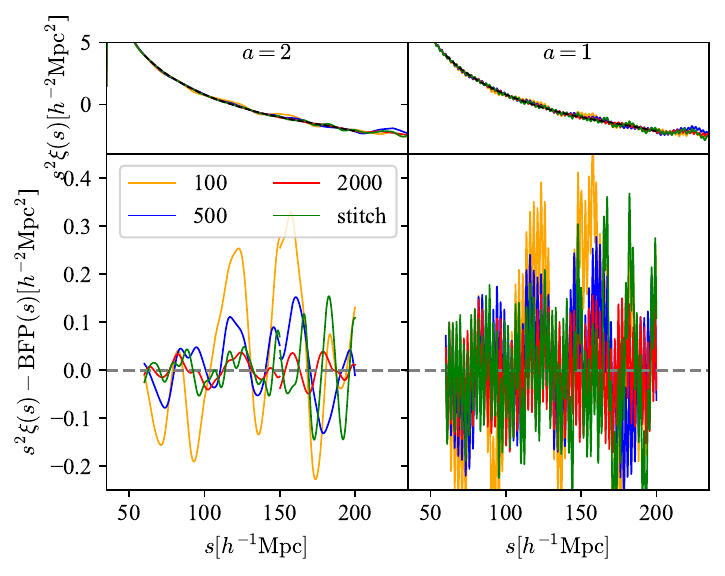}
 \caption{Upper panels: coloured curves - The result of the transformation expressed by Eq.~\eqref{eq:dampfouriertransform} of the SICKLE power spectra computed as the average of 100, 500, 2000 realisations and by stitching the average of 2000 realisations with the one of 50 realisations (read text for details); black curve - the best-fitting polynomial of the red curve. Lower panels: the difference between the upper coloured curves and the black curve. The left and right panels correspond to a different damping parameter (Eq.~\eqref{eq:dampfouriertransform}), i.e. $a=2$ and $a=1$, respectively.}
 \label{fig:SK_noise_B_a1a2}
\end{figure}

In Figure~\ref{fig:SK_noise_B_a1a2}, one can see the 2PCF computed as the Hankel transform of the average void model power spectrum, for two different damping factors ($a=1~h^{-1}\mathrm{Mpc}$ and $a=2~h^{-1}\mathrm{Mpc}$). 
The black curves in the upper panels represent best-fitting polynomials (BFP) of the $s^2\xi(s)$ curve -- computed using Eq.~\eqref{eq:dampfouriertransform} and the average of 2000 power spectrum realisations -- for two different $s$ intervals: $s\in(60, 150)\,h^{-1}\mathrm{Mpc}$ and $s\in(150, 200)\,h^{-1}\mathrm{Mpc}$.
The lower panels of Figure~\ref{fig:SK_noise_B_a1a2} contain the differences between $s^2\xi(s)$ curves and the BFP. 

Apart from the visual inspection of the noise in the 2PCF, we also quantify it by computing:
\begin{equation}
\label{eq:BFP_noise_a1a2}
    \Phi = \frac{1}{n}\sum_{i=1}^n \left[ s_i^2 \xi(s_i) - \mathrm{BFP}(s_i) \right]^2,
\end{equation}
where $n$ is the number of bins in the given interval and $i$ is the index of the bin.
One can observe from Figure~\ref{fig:SK_noise_B_a1a2} and Table~\ref{tab:bfp_noise_a1_a2} that the noise is drastically reduced when the number of realisations is increased from 100 to 2000.

As mentioned in Section~\ref{sec:power_spectrum}, we need:
\begin{itemize}
    \item a grid size of $2048^3$ to measure the power spectrum for a large enough $k$ interval,
    \item a large number of realisations to minimise the effect of the noise (cosmic-variance),   
\end{itemize}
but achieving both conditions simultaneously is computationally-expensive. 
Thus, we create a stitched model by computing 2000 power spectra using a grid size of $512^3$ (to decrease the noise at large scales) and 50 power spectra using a grid size of $2048^3$ (to have a reasonably de-noised power spectrum up to a large value of $k$). Figure~\ref{fig:SK_noise_B_a1a2} and Table~\ref{tab:bfp_noise_a1_a2} suggest that the stitched model performs at least as well as the 500 case for the really large scales and reaches the precision of the 1000 case for the lower scales.

\begin{figure}
 \includegraphics[width=\columnwidth]{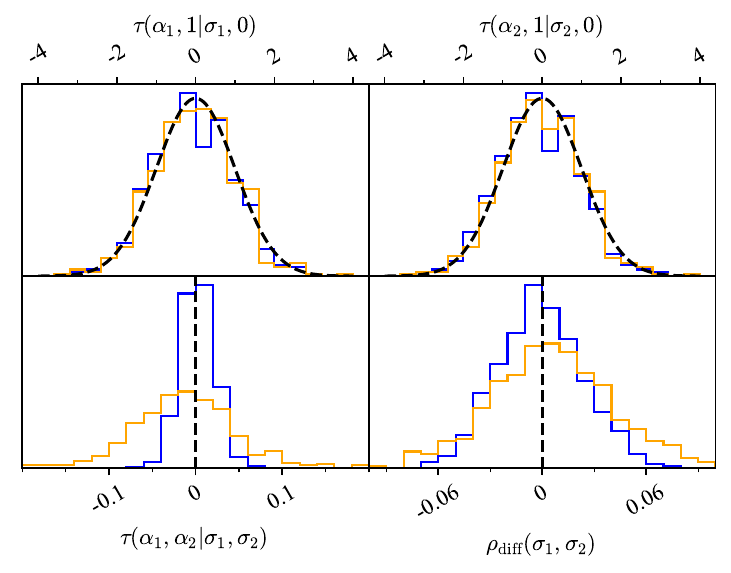}
 \caption{Upper panels: $(\alpha-1)/\sigma$ for $a=1~h^{-1}\mathrm{Mpc}$ ($\alpha_1$, $\sigma_1$, left) and $a=2~h^{-1}\mathrm{Mpc}$ ($\alpha_2$, $\sigma_2$, right). Lower panels: the tension between the $\alpha$ values measured using $a=1~h^{-1}\mathrm{Mpc}$ and $a=2~h^{-1}\mathrm{Mpc}$ (left) and the relative difference between the uncertainties ($\sigma$) on $\alpha$ (right). Blue histograms: the results for the parabolic model with a PAR$_\mathrm{G}$ prior on $c$. Orange histograms: the results for the CG$_\mathrm{B}$ numerical model. $a$ is the damping parameter from Eq.~\eqref{eq:dampfouriertransform}.  The histograms contain the results of 500 individual 2PCF computed from \textsc{Patchy} cubic mocks. }
 \label{fig:damp_parameter_a1a2_alpha_sigma}
\end{figure}

One can also observe in Figure~\ref{fig:SK_noise_B_a1a2} and Table~\ref{tab:bfp_noise_a1_a2} that the damping factor $a$ impacts the noise levels. By increasing it from $a=1~h^{-1}\mathrm{Mpc}$ to $a=2~h^{-1}\mathrm{Mpc}$ the amplitude of the noise is reduced by almost one order of magnitude. Consequently, we have tested whether the value of $a$ can bias the measurement of $\alpha$, by computing the tensions Eq.\,\eqref{eq:tension_parameter} between the $\alpha$ values corresponding to $a=1~h^{-1}\mathrm{Mpc}$ ($\alpha_1$) and $a=2~h^{-1}\mathrm{Mpc}$ ($\alpha_2$). Figure~\ref{fig:damp_parameter_a1a2_alpha_sigma} shows that there is no tension between the two cases and for both cases, the histogram of the 500 $\tau(\alpha, 1| \sigma, 0)$ values are consistent with a standard-normal distribution, meaning there is no bias and the uncertainties are correctly estimated. Moreover, the relative difference $\rho_\mathrm{diff}= (\sigma_1 - \sigma_2) / [0.5\times(\sigma_1 + \sigma_2)]$ shows that there is no bias in the uncertainty estimation between the two cases.
Given the previous reasons, we fix $a=2~h^{-1}\mathrm{Mpc}$ in the current paper.

\begin{table}
    \centering
    \begin{tabular}{c|c|c}
        \hline
             $\Phi$ for & 60--150  & 150--200 \\
             $a=1~h^{-1}\mathrm{Mpc}$ & $\times 10^{-3}$ & $\times 10^{-3}$ \\
        \hline
            100    & $30.0$ & $44.7$ \\
            200    & $14.6$ & $35.4$ \\
            500    & $10.2$ & $14.9$ \\
            1000   & $7.60$ & $9.60$ \\
            2000   & $6.60$ & $7.14$ \\
            stitch & $10.3$ & $27.2$ \\
        \hline
            $\Phi$ for               & 60--150          & 150--200\\
            $a=2~h^{-1}\mathrm{Mpc}$ & $\times 10^{-4}$ & $\times 10^{-4}$ \\
        \hline
            100    & $189.0$ & $306.0$ \\
            200    & $50.0$ & $235.0$ \\
            500    & $29.5$ & $73.1$ \\
            1000   & $9.48$ & $23.9$ \\
            2000   & $3.56$ & $6.16$ \\
            stitch & $8.66$ & $74.5$ \\
        \hline
    \end{tabular}
    \caption{The $\Phi$ values defined in Eq.~\eqref{eq:BFP_noise_a1a2} for two $s$ intervals $s\in(60, 150)~h^{-1}\mathrm{Mpc}$ and $s\in(150, 200)~h^{-1}\mathrm{Mpc}$ and for two values of the damping factor $a=1~h^{-1}\mathrm{Mpc}$ and $a=2~h^{-1}\mathrm{Mpc}$. \label{tab:bfp_noise_a1_a2}}
\end{table}

\begin{figure}
 \includegraphics[width=\columnwidth]{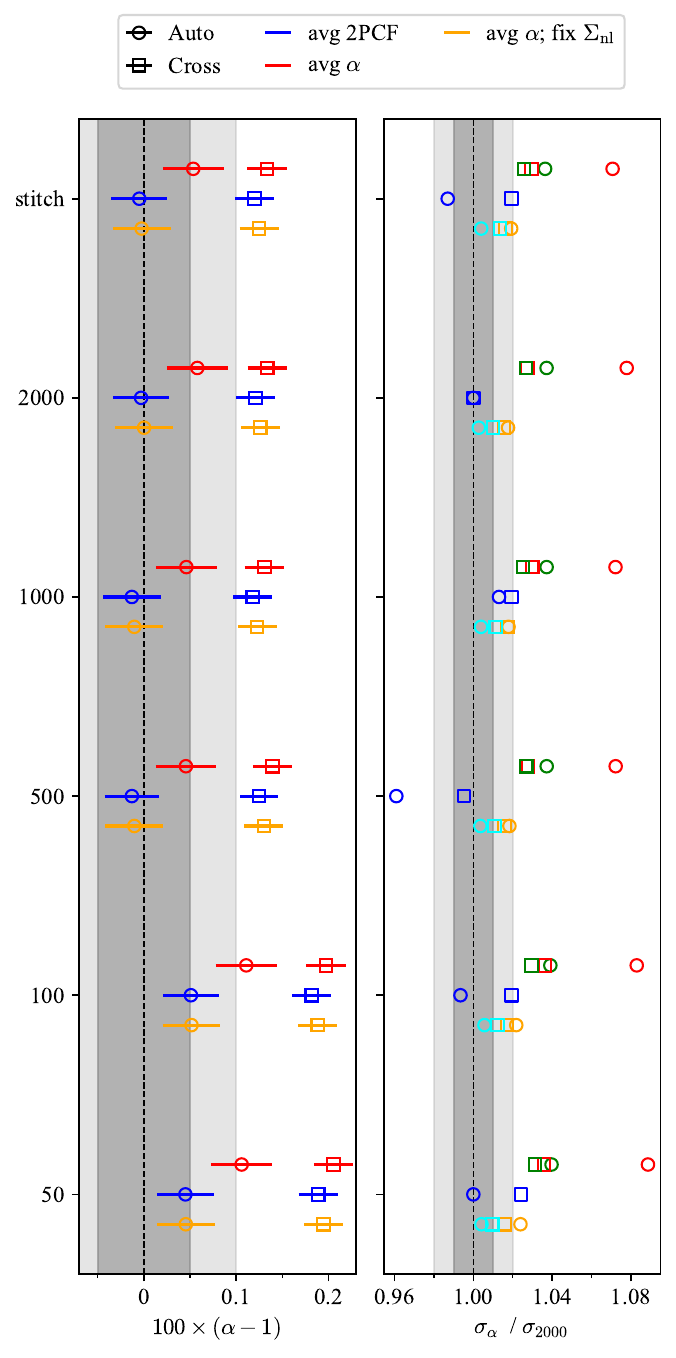}
 \caption{Comparison between the results of the model power spectra (SICKLE) computed from different number of realisations -- 50, 100, 500, 1000, 2000 -- and by stitching (see text), using the void auto- and void-galaxy cross-2PCF computed from 500 pre-reconstructed \textsc{Patchy} cubic mocks. First column shows the bias of $\alpha$ with respect one. The second column contains the ratios between different uncertainty estimations and the blue coloured $\sigma_\mathrm{2000}$. The three colours denote the ways the fitting has been performed: blue - on the average of the 500 2PCF, with a rescaled covariance matrix (by 500), thus $\alpha$ is the median of the posterior distribution and $\sigma_\alpha$ is half the difference between the 84th and 16th percentiles; red and green - on the individual 2PCF, with the normal covariance matrix; orange and cyan - similarly to red and green, but with a fixed $\Sigma_\mathrm{nl}$. For red and orange, the shown $\alpha$ and $\sigma_\alpha$ are the average and the standard deviation -- divided by $\sqrt{500}$ -- of the 500 $\alpha$ values, respectively. For green and cyan, $\sigma_\alpha$ is the mean of the 500 $\sigma$ provided by the individual fittings of the 2PCF, divided by the $\sqrt{500}$}
 \label{fig:A_alpha_avg_SK_realisations}
\end{figure}

After fixing $a=2~h^{-1}\mathrm{Mpc}$, we also test whether different number of realisations for the model power spectra and the stitching method affect the $\alpha$ measurements and the corresponding uncertainties. Figure~\ref{fig:A_alpha_avg_SK_realisations} shows a comparison between the results of the model power spectra (SICKLE) computed from different number of realisations -- 50, 100, 500, 1000, 2000 -- and by stitching.
We study three fitting scenarios:
\begin{enumerate}
    \item on the average of the 500 \textsc{Patchy} 2PCF with a rescaled covariance matrix (blue);
    \item on the individual 2PCF, with the normal covariance matrix (red and green);
    \item on the individual 2PCF, with the normal covariance matrix, but with a fixed $\Sigma_\mathrm{nl}$ (orange and cyan).

\end{enumerate}
The shown $\alpha$ and $\sigma_\alpha$ corresponding to the three previous cases are, respectively: (i) the median of the posterior distribution and half the difference between the 84th and 16th percentiles;
(ii) and (iii) the average and the standard deviation -- divided by $\sqrt{500}$ -- of the 500 $\alpha$ values (red and orange). Additionally, the cyan and the green points denote the mean of the 500 $\sigma$ provided by the individual fittings of the 2PCF, divided by the $\sqrt{500}$.
The uncertainties on the right panel from void auto-2PCF and void-galaxy cross-2PCF are divided by the corresponding blue $\sigma_\mathrm{2000}$, which explains why the blue square and circle for the 2000 case are exactly positioned at one.

On one side, one can observe that starting from the '500' model, the $\alpha$ converges to the same value, for both void auto- and void-galaxy cross-2PCF and in all three fitting scenarios. On the other side, all the ways to estimate the uncertainty provide $\sigma_\alpha$ values that are consistent within one to two per cent between all models and per method, except the '500' void auto-2PCF blue case, where the deviation is around four per cent.
Consequently, the stitched method is chosen as the standard way to construct the void model power spectrum throughout this paper.

We also fit the individual 2PCF with a fixed  $\Sigma_\mathrm{nl}$ -- in Figure~\ref{fig:A_alpha_avg_SK_realisations} because we have observed that the noise in the \textsc{Patchy} void 2PCF allows for larger values of $\Sigma_\mathrm{nl}$ to fit the data, which enlarges the posterior of $\alpha$ towards larger values. This slightly biases the measurement and overestimates the uncertainty. Given that throughout the paper we have not fixed $\Sigma_\mathrm{nl}$ for boxes, one has to consider this 0.05 per cent bias in the results of the main text.

\section{The study of the nuisance parameters}
\label{sec:appendix_nuisance_parameters}

Given the fact that the Least-Squares (LS) is much faster than \textsc{PyMultiNest}, in the main analysis, we use a two--fold approach in order to reduce the fitting time:
\begin{itemize}
    \item \textsc{PyMultiNest} to fit $\alpha$, $B$, $\Sigma_\mathrm{nl}$, $c$;
    \item LS to fit the nuisance parameters $a_0$, $a_1$, $a_2$.
\end{itemize}
In this section, we show that this approach does not bias the measurements of $\alpha$, $B$, $\Sigma_\mathrm{nl}$, $c$ and that there are no degeneracies between the nuisance parameters and $\alpha$. To verify this, we fit the average void auto-2PCF and the average void-galaxy cross-2PCF computed from 500 pre-reconstructed \textsc{Patchy} cubic mocks, using a rescaled covariance matrix (i.e. divided by 500). Given that DW is not performing well, we only test the CG$_\mathrm{B}$, SK$_\mathrm{B}$ and PAR$_\mathrm{U}$ models.

\begin{figure}
 \includegraphics[width=\columnwidth]{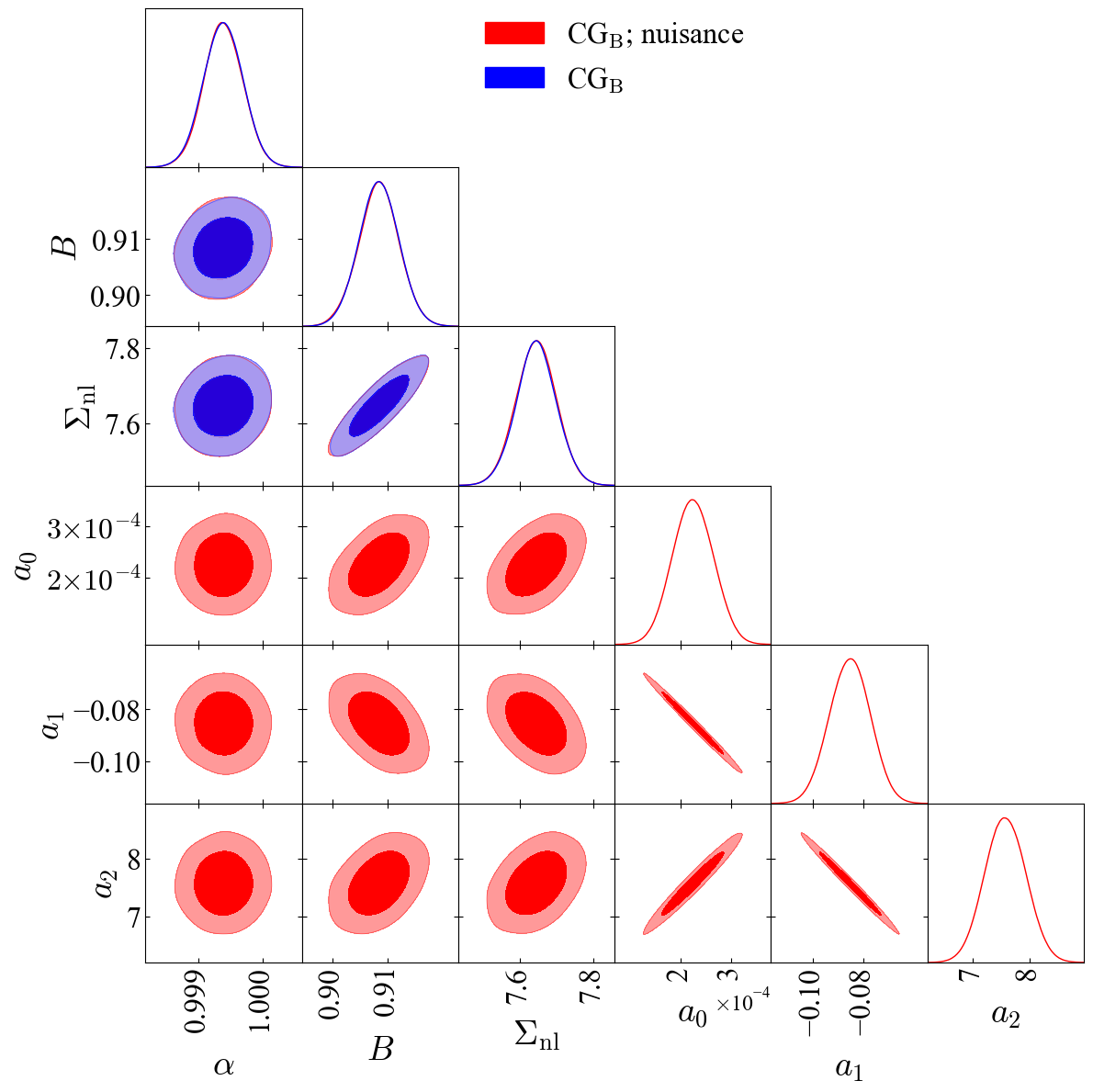}
 \caption{Triangle plot containing the posterior distributions of the fitting parameters described in Section~\ref{sec:baomodel}. The fitting has been performed on the average void auto-2PCF computed from 500 pre-reconstructed \textsc{Patchy} cubic mocks using the CG$_\mathrm{B}$ numerical model. Red - all six parameters are given to \textsc{PyMultiNest}; Blue - only $\alpha$, $B$ and $\Sigma_\mathrm{nl}$ are given to \textsc{PyMultiNest}, while the nuisance parameters are fitted using a Least-Square method.}
 \label{fig:BAOfit_avg_16R_500_2pcf_cg}
\end{figure}

\begin{figure}
 \includegraphics[width=\columnwidth]{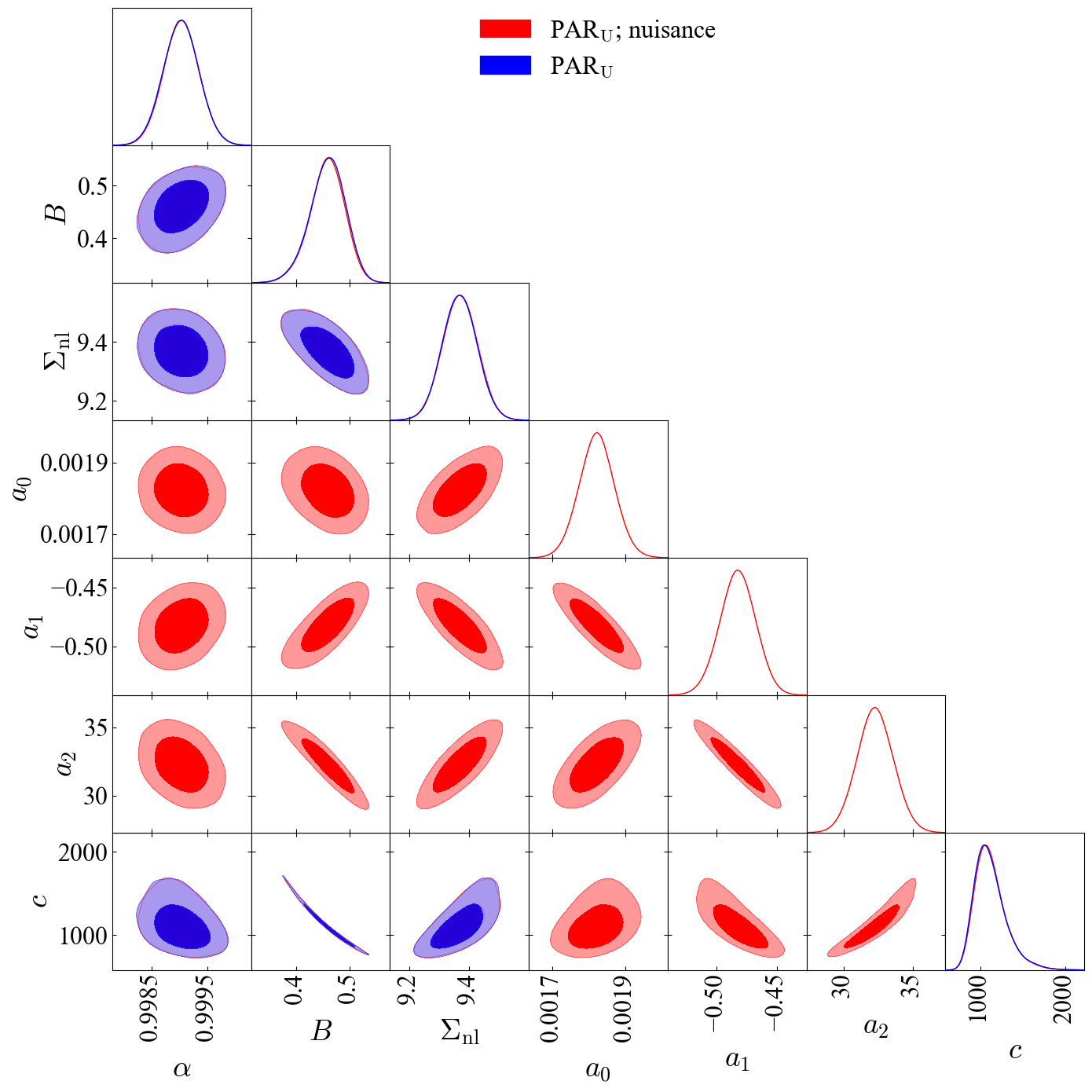}
 \caption{Same as Figure~\ref{fig:BAOfit_avg_16R_500_2pcf_cg}, but the model is PAR$_\mathrm{U}$.}
 \label{fig:BAOfit_avg_16R_500_2pcf_par}
\end{figure}

\begin{figure}
 \includegraphics[width=\columnwidth]{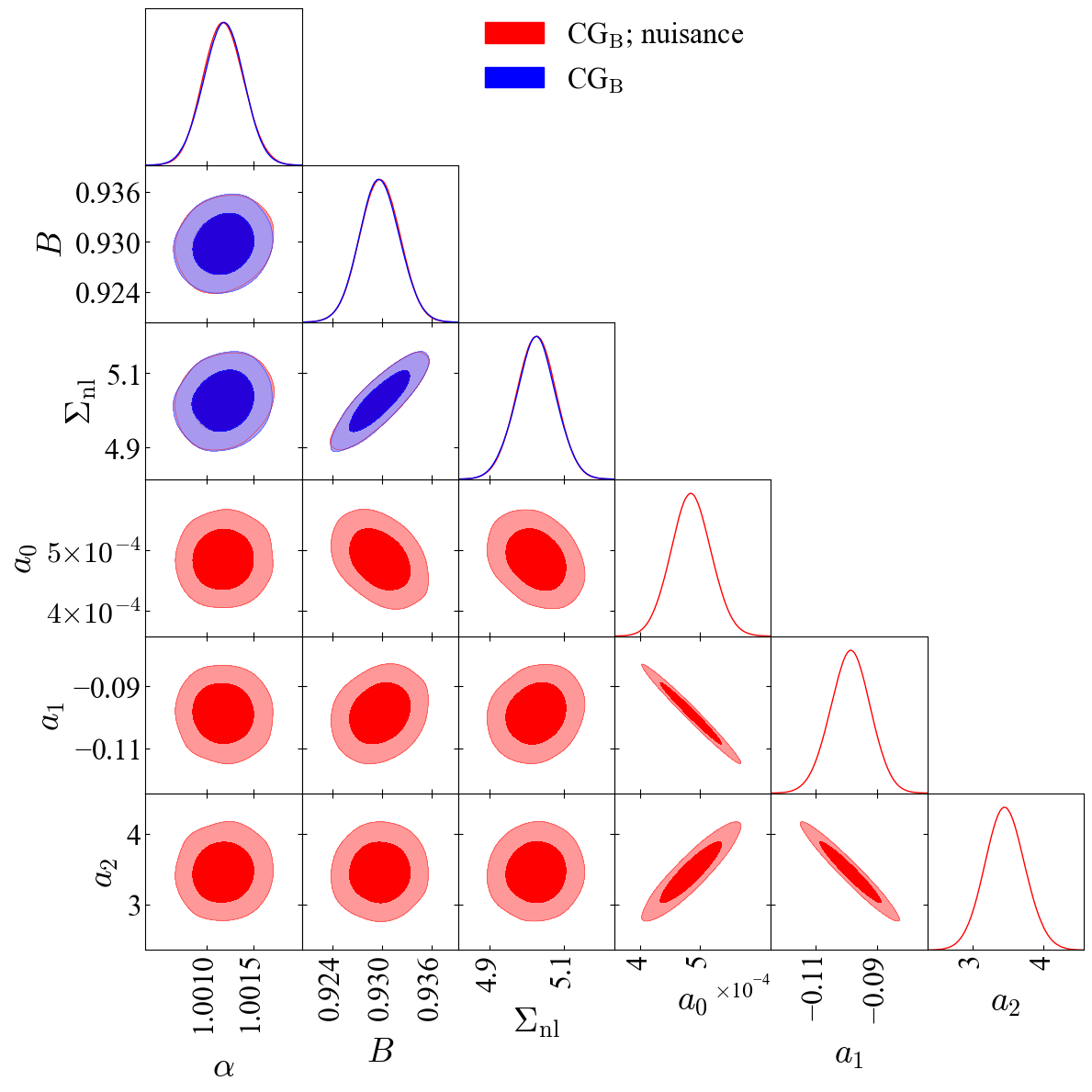}
 \caption{Same as Figure~\ref{fig:BAOfit_avg_16R_500_2pcf_cg}, but the reference is the average void-galaxy cross-2PCF.}
 \label{fig:BAOfit_avg_16R_500_2pcf_xcg}
\end{figure}

\begin{figure}
 \includegraphics[width=\columnwidth]{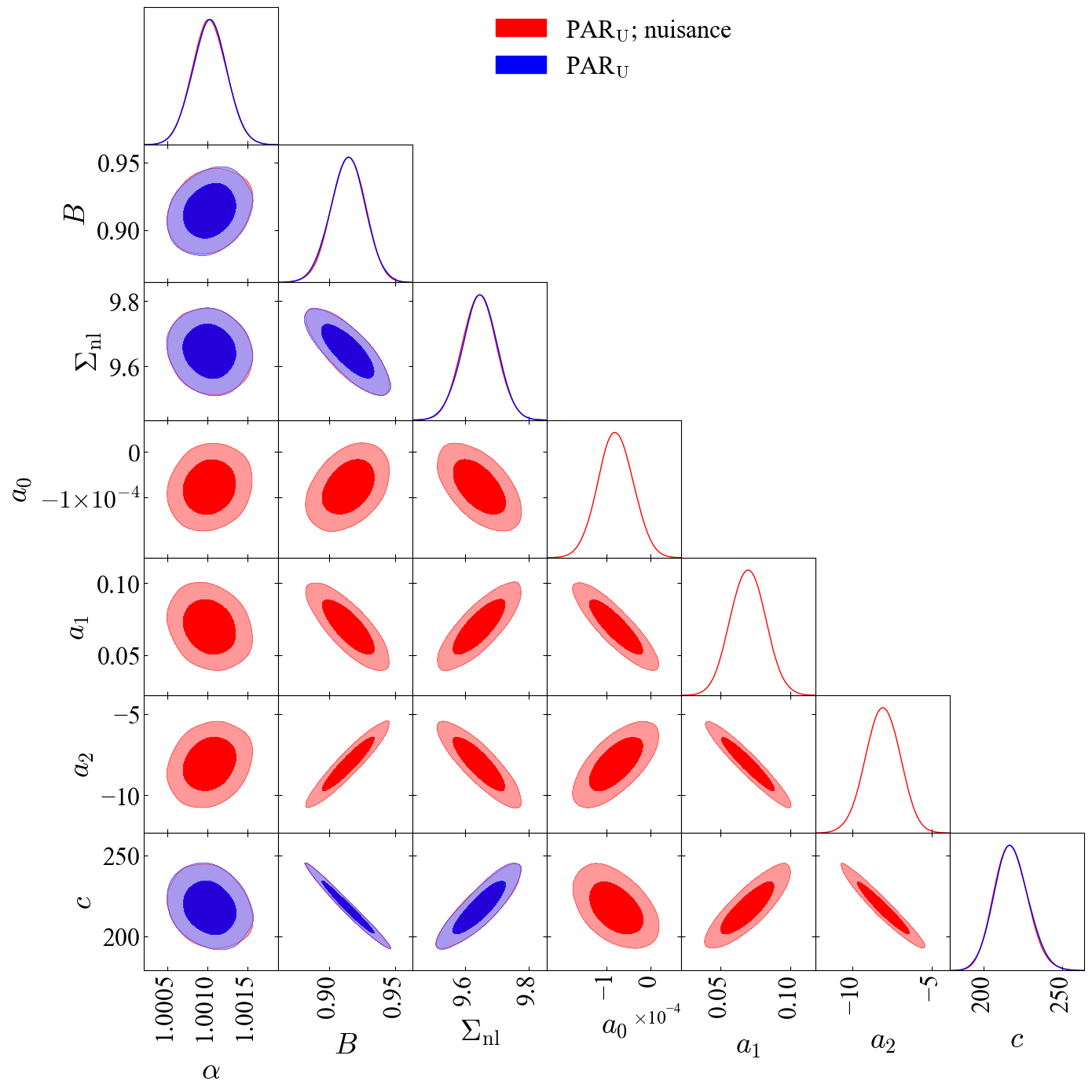}
 \caption{Same as Figure~\ref{fig:BAOfit_avg_16R_500_2pcf_cg}, but the model is PAR$_\mathrm{U}$ and the reference is the average void-galaxy cross-2PCF.}
 \label{fig:BAOfit_avg_16R_500_2pcf_xpar}
\end{figure}

Looking at the best-fitting nuisance parameters in Table~\ref{tab:nuisance_params}, SK$_\mathrm{B}$ behaves similarly to CG$_\mathrm{B}$, thus we further focus on CG$_\mathrm{B}$ and PAR$_\mathrm{U}$. Figures~\ref{fig:BAOfit_avg_16R_500_2pcf_cg}, \ref{fig:BAOfit_avg_16R_500_2pcf_par}, \ref{fig:BAOfit_avg_16R_500_2pcf_xcg} and \ref{fig:BAOfit_avg_16R_500_2pcf_xpar} show the posterior distributions of the fitting parameters in two cases: red -- all six or seven parameters are sampled by \textsc{PyMultiNest}; blue -- the two--fold approach. In the first case, we used the following priors for the nuisance parameters: $p(a_0) = \mathcal{U}_{[-1, 1]}(a_0)$, $p(a_1) = \mathcal{U}_{[-10, 10]}(a_1)$ and $p(a_2) = \mathcal{U}_{[-100, 100]}(a_2)$, that are wide enough to not influence the fitting results.


\begin{table}
    \centering
    \begin{tabular}{c|c|c|c}
        \hline
             & $a_0 [10^{-4}]$ & $a_1 [10^{-2} \times h^{-1}\mathrm{Mpc}]$ & $a_2 [h^{-2}\mathrm{Mpc}^2]$ \\
        \hline
            CG$_\mathrm{B}$ & $2.3$ ($4.9$) & $-8.7$ ($10$) & 7.6 (3.5) \\
            SK$_\mathrm{B}$ & $5.4$ ($4.8$) & $-12 $ ($-8.6$) & 7.4 (2.5)\\
            PAR$_\mathrm{U}$ & $18$ ($0.69$) & $-47$ ($6.7$)  & 31 (-7.9) \\
        \hline
    \end{tabular}
    \caption{The best-fitting nuisance parameters for three models. The fitting has been performed on the average void auto-2PCF and void-galaxy cross-2PCF (in brackets) computed from 500 pre-reconstructed \textsc{Patchy} boxes. The abbreviations are defined in Table~\ref{tab:model_abbreviations}.\label{tab:nuisance_params}}
\end{table}

The same figures reveal that the measurements of $\alpha$, $B$, $\Sigma_\mathrm{nl}$ and $c$ are insensitive to the inclusion of the nuisance parameters in the \textsc{PyMultiNest} chain as the blue curves are consistent with the red ones. In the PAR$_\mathrm{U}$ case, there are slight degeneracies between $\alpha$ and $a_1$, $a_2$, however, they may be caused by the introduction of the $c$ parameter and its strong degeneracy with $a_1$, $a_2$. In contrast, for CG$_\mathrm{B}$, $\alpha$ is not degenerate with the nuisance parameters. These results are consistent with the observations provided by \citet{ImproveBAOvoids_BOSS, 10.1093/mnras/stac390} and with the fact that the nuisance parameters should describe the broad-band shape.

Consequently, we argue that one can safely use the combined \textsc{PyMultiNest} -- LS approach in order to measure the fitting parameters.

\section{Light-cone results}
\label{sec:light-cone_results}

As mentioned in Section~\ref{sec:survey-geometry_effects}, we have only shown the results for \textsc{CosmoGAME} in the main text due to visibility reasons. Here, we show a comparison between all models $\mathrm{CG}_\mathrm{B}$, $\mathrm{SK}_\mathrm{B}$, $\mathrm{CG}_\mathrm{LC}$, $\mathrm{SK}_\mathrm{LC}$ and parabolic model with fixed $c$.

\begin{figure}
 \includegraphics[width=\columnwidth]{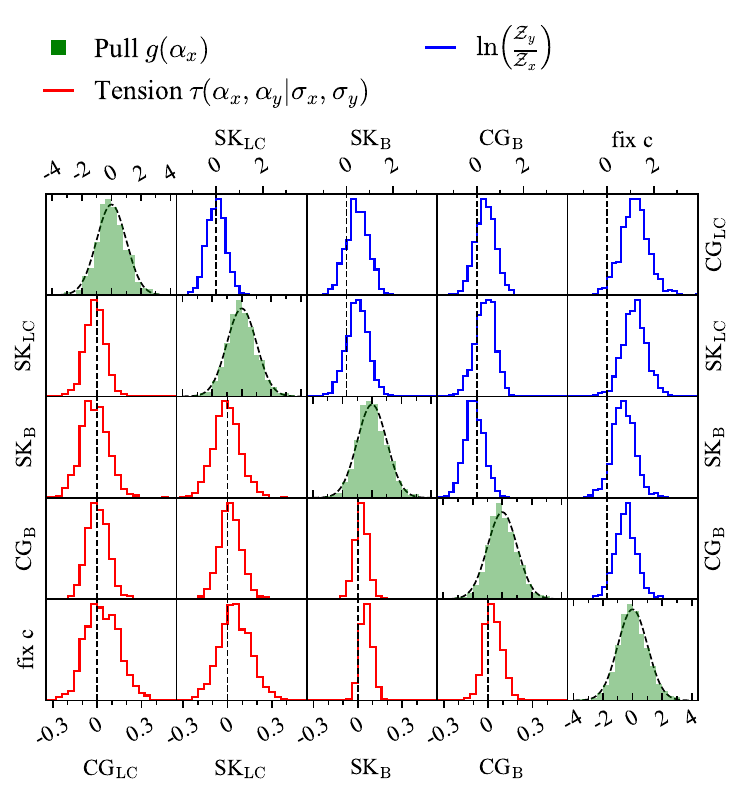}
 \caption{Diagonal panels: green - the histograms of the pull function $g(\alpha_x)$ values, Eq.~\eqref{eq:pull_function}; black - standard normal distributions. Lower triangular plots: the values of $\tau(\alpha_x, \alpha_y| \sigma_x, \sigma_y)$, Eq.~\eqref{eq:tension_parameter}, for all combinations of models. Upper triangular plot: the natural logarithm of the Bayes Factor $\ln{\left(\mathcal{Z}_y/\mathcal{Z}_x\right)}$ (see Section~\ref{sec:bayes_fac}). The results correspond to the individual fittings of the 1000 void auto-2PCF computed from the \textsc{Patchy} light-cone mocks. The abbreviations are defined in Table~\ref{tab:model_abbreviations}.}
 \label{fig:LC_evi_tension_pullone_auto_all}
\end{figure}

\begin{figure}
 \includegraphics[width=\columnwidth]{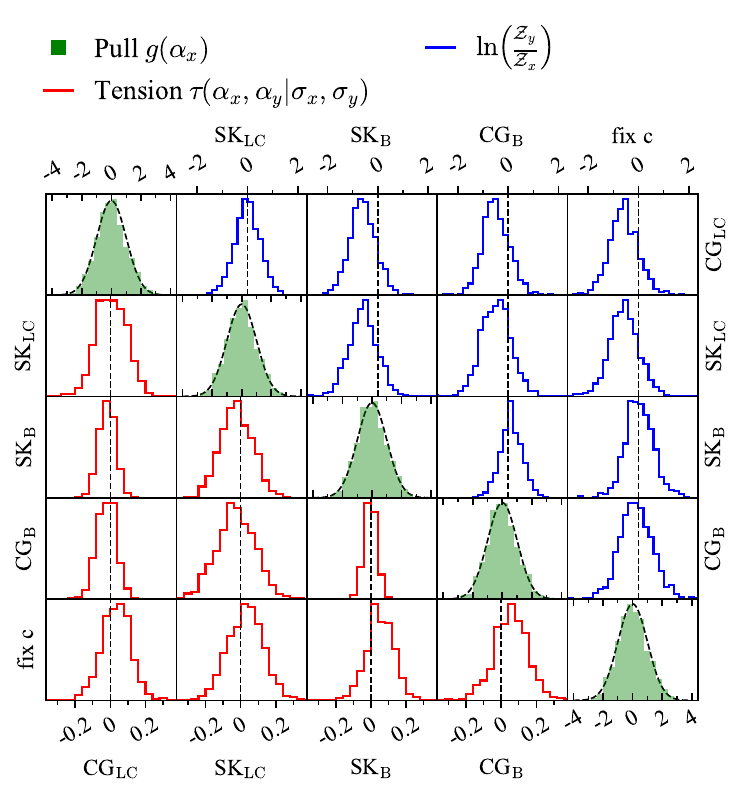}
 \caption{Same as Figure~\ref{fig:LC_evi_tension_pullone_auto_all}, but for void-galaxy cross-2PCF.}
 \label{fig:LC_evi_tension_pullone_cross_all}
\end{figure}

Studying the tension in the lower diagonal plots of Figures~\ref{fig:LC_evi_tension_pullone_auto_all} and \ref{fig:LC_evi_tension_pullone_cross_all}, we observe that the box-based models and the light-cone based models provide highly consistent results. There is however a slight bias of the order of $0.1\sigma$ between the fixed $c$ parabola and the numerical models. All models estimate accurately the uncertainty of $\alpha$. The logarithm of the Bayes factor suggests that for the void auto-2PCF, the fixed $c$ parabola is slightly disfavoured against the numerical models, while for the void-galaxy cross-2PCF, the reverse is true. Moreover, there are no siginificant differences between CG$_\mathrm{B}$ and SK$_\mathrm{B}$, nor between CG$_\mathrm{LC}$ and SK$_\mathrm{LC}$.
Lastly, for the void auto-2PCF, the light-cone numerical models are slightly preferred compared to the ones constructed for boxes, while the opposite is valid for the void-galaxy cross-2PCF.

